\documentclass[12pt]{article}
\usepackage{amssymb, amsmath, amsthm, physics, graphicx, hyperref, xcolor,amsfonts} 
\usepackage[margin = 0.5in]{geometry}
\usepackage{bm}
\usepackage{cancel}

\numberwithin{equation}{subsection}
\newcommand{\I}{\mathbb{I}}

\begin{document}
\include{front}

\thispagestyle{empty}
\noindent{\small
\hfill{$~~$}  \\ 
{}
}
\begin{center}
{\large \bf
10D Supergravity Numerical Data Sets for L $\&$ R Matrices \\
\vskip4pt
$~~$
}  \\   [8mm]
{\large {
Jacob Cigliano\footnote{cigliano@terpmail.umd.edu}${}^{, a}$,
Bergen Dahl \footnote{bdahl@terpmail.umd.edu}${}^{,a}$, and
S. James Gates, Jr. \footnote{gatess@umd.edu}${}^{,b,c}$
}}
\\*[6mm]
\emph{
\centering
$^{a}$University of Maryland, Department of Physics
\\[1pt]
College Park, MD 20742
\\[10pt]
}
\emph{
$^{b}$University of Maryland, Department of Physics
\\[1pt]
J. S. Toll Hall, Room 1117, College Park, MD 20742, USA
\\[10pt]
}
\emph{
$^{c}$University of Maryland, School of Public Policy
\\[1pt]
T. Marshall Hall, Room 2223, College Park, MD 20742, USA
\\[10pt]
}
\vspace*{80mm}
{ ABSTRACT}\\[05mm]
\parbox{142mm}{\parindent=2pc\indent\baselineskip=14pt plus1pt
After reviewing the development of 10D, superspace theories, and their relations to superstring and heterotic string theories, explicit calculations are undertaken in the on-shell $\mathcal N = 1$ linearized supergravity, the
associated supercurrent is derived, and non-closure terms are explicitly given. The $L_I$ and $R_I$ adjacency matrices are then computed in complete numerical form as data sets.  This is the preliminary step required to perform a scan to embed the
on-shell matrices into off-shell ones.
} \end{center}

\newpage
\tableofcontents

\newpage
\section{Introduction}
\indent
\\
Attention to the realization of supersymmetry in Minkowski space with one temporal and nine spatial dimensions began to emerge in
the 1970's and extended into the 1980's. \cite{GSO1976, GSO1977, BrinkSchwarzScherk1977, Chamseddine1981, Nilsson1981,GreenSchwarz1981, BergshoeffDeRoo1982, BergshoeffEtAl1982, GreenSchwarz1982, ChaplineManton1983, GreenSchwarz1984}
It was critical to the development of first superstring theory and later heterotic string theory \cite{Goddard:1985,Gross:1985PRL,Gross:1985NPB256,Gross:1986NPB267,Sen:1985PRL,Sen:1985PRD,CHSW:1985NPB,Narain:1986PLB,NarainSarmadiWitten:1987NPB,AlvarezGaume:1986PLB,DixonHarvey:1986NPB}. One of the authors (SJG) during
the 1980's collaboratively worked on a number of such explorations in the works seen in \cite{GatesNishino:1985,GatesNishino:1986a,GatesNishino:1986b,NishinoGates:1987Euler,GatesVashakidze:1987I,GatesNishino:1987II,CarrGatesOerter:1987,BellucciGates:1988}.
\newline

\noindent\textbf{Citation Trajectory \#1} shows the clean progression from recognizing supersymmetric strings' low-energy content, to identifying the privileged higher-dimensional gauge theories, to constructing the explicit interacting supergravity--Yang--Mills system in $D{=}10$, and finally to proving its anomaly freedom. The arc is the maturation of ten-dimensional $N{=}1$ supergravity as the effective backbone of superstring theory.
\newline $~$ 

\noindent\textbf{Citation Trajectory \#2} gives a rapid progression from \emph{constructing} a new anomaly-free superstring (heterotic) and its gauge groups, to \emph{cementing} its world-sheet realization and deriving its effective spacetime equations, to \emph{exploring} compactifications and dualities via lattice methods, while also \emph{testing the bounds of consistency} through non-supersymmetric ten-dimensional vacua. Collectively, these papers establish the heterotic string as a technically complete theory with rich moduli spaces and a broader landscape than supersymmetric models alone.
\newline

\noindent\textbf{Citation Trajectory \#3} demonstrates progress from proposing a dual, anomaly-ready $D{=}10$, $N{=}1$ superspace supergravity, to proving its compatibility with Green--Schwarz superstrings, to computing $O(\alpha')$ corrections in a manifestly supersymmetric way, and then to expanding the formalism to include topological invariants, higher-order effects, and Type~II superspace supergravities. Collectively the papers establish a robust superspace infrastructure for ten-dimensional string-effective supergravity.

\section{Motivation}
\indent
In a series of works that began in the early 2000's, a radical reformulation was proposed with the introduction
of the idea that supersymmetrical theories could be reformulated in terms of a set of graphs given the
names of ``adinkras,'' \cite{FG:2005,adnk2,adnk3,adnk4,Gates:2009G-1,Gates:2012G-2,Gates:2013G-3,Gates:2013G-4}
These works first introduced the mathematical properties of these graphs, and then demonstrated
how they are obtained from 4D, $\cal N$ = 1 supermultiplets containing particles with spins from
zero to two as well as the special case of 4D, $\cal N$ = 4 SYM theory.

In a series of works during the 2020's \cite{Gates:2019Codex,Gates:2020Prepot10D,Gates:2021Weyl10D} there
was a returned focus to SG theories in 10D and 11D, with the emphasis on laying a foundation
for extending the work completed in 4D to these higher dimensional theories.  The efforts completed
in these works included:
\begin{itemize}
 \item Construction of superspace formalism aimed at prepotential versions of $11$D and $10$D supergravity, explicitly motivated by the absence of known full off-shell theories, supplying extra structure needed for later off-shell searches.

\item
With the scalar-superspace laboratory in place, the 2020 JHEP paper performs complete Lorentz decompositions of unconstrained $10$D scalar superfields and introduces $10$D Adinkras. Branching-rule technology converts superfield expansions into representation-theoretic data, enabling a systematic scan for superfields containing graviton/gravitino components---precisely the data required to identify off-shell prepotential candidates.

\item
Finally, the Weyl-covariance paper supplies the superconformal transformation laws and Weyl field strengths in $10$D superspace, and uses them to propose a finite list of $\mathcal{N}{=}1$ $10$D prepotential superfields. This closes the loop: the Nordstr\"om laboratory motivates the search, Adinkra/adynkra methods make component content explicit, and Weyl covariance provides the organizing principle for off-shell/ superconformal prepotentials.

\end{itemize}

 A primary purpose of the current work is to obtain the required matrix input from 10D SG investigations completed in the 2020's required to leverage these new tools developed in 2000's.

\section{Superalgebra}
\subsection{Lagrangian and Equations of Motion}
The on-shell 10D, $\mathcal N = 1$ supergravity multiplet consists of a symmetric tensor $h_{\mu \nu}$ (the graviton), a vector-spinor $\psi_{\mu a}$ (the gravitino), a two-form $B_{\mu \nu}$, and the scalar spinor pair $\phi, \chi_a$ (the dilaton and dilatino). Imposing Lorentz invariance, reality, and engineering-dimension constraints, the most general quadratic free-field Lagrangian built from these fields takes the form
\begin{equation}
    \mathcal L = -2 \mathcal{R}_0 + 2i \psi_{\mu}{}^a (\sigma^{\mu\nu\rho})_a{}^{\dot b} \partial_{\nu} \psi_{\rho \dot b} -\frac{3}{2} A_{[3]}A^{[3]} -4i  \chi^{\dot c}(\tilde{\sigma}^\mu)_{ \dot c}{}^{b} \partial_{\mu} \chi_b -\frac{1}{2}\partial_{\mu} \phi \partial^{\mu} \phi
\end{equation}
where the coefficients have been fixed uniquely by supersymmetry closure and supercurrent normalization. The three-form field strength tensor is defined by $A_{\mu \nu \rho} = \frac{1}{3!}\partial_{[\mu}B_{\nu \rho]}$, and the linearized Einstein term is
\begin{equation}\mathcal{R}_0 =  \frac{1}{4}\partial_{\alpha}h_{\mu\nu}\partial^{\alpha}h^{\mu\nu} -\frac{1}{4}\partial^\alpha h\partial_\alpha h +\frac{1}{2}\partial^\alpha h \partial^\beta h_{\alpha\beta} - \frac{1}{2}\partial^{\mu} h_{\mu \nu} \partial_\alpha h^{\alpha \nu}\end{equation}
with $h = h_{\nu}{}^\nu$. In this paper, index (anti)symmetrization brackets are unnormalized: for example,
\[A_{[\mu}B_{\nu]} = A_\mu B_\nu - A_\nu B_\mu.\]
The term $\mathcal R_0$ is the standard Fierz--Pauli Lagrangian for a linearized massless spin-2 field.
Its structure ensures invariance under linearized diffeomorphisms, $\delta h_{\mu\nu}=\partial_{(\mu}\xi_{\nu)}$. Similarly, the two-form enters only through its field strength $A_{\mu\nu\rho}$, reflecting the Abelian gauge symmetry $\delta B_{\mu\nu}=2\partial_{[\mu}\Lambda_{\nu]}$. Together with the dilaton $\phi$, this bosonic sector matches the familiar NS--NS field content in ten-dimensional supergravity.

Varying the resulting action yields the ten-dimensional analogs of the Rarita-Schwinger and Dirac equations, respectively.
\begin{align}
    &(\sigma_{\mu}{}^{\nu \rho})_a{}^{\dot b} \partial_\nu \psi_{\rho \dot b} = 0\\&\cancel \partial_{\dot a}{}^{b}\chi_b = 0
\end{align}
The gravitino equation admits several equivalent forms corresponding to its $\tilde \sigma$-trace and to gauge-fixed projection. Left multiplying by a contracted $\tilde \sigma^{[1]}$ gives
\begin{align}
    &(\tilde \sigma^\mu)_{\dot a}{}^c (\sigma_{\mu}{}^{\nu \rho})_c{}^{\dot b} \partial_\nu \psi_{\rho \dot b} = 8(\tilde \sigma^{\nu \rho})_{\dot a}{}^{\dot b} \partial_\nu \psi_{\rho \dot b}\tag*{}\\
    \implies \quad & (\tilde \sigma^{\nu \rho})_{\dot a}{}^{\dot b} \partial_\nu \psi_{\rho \dot b} = 0
\end{align}
Decomposing the Clifford algebra element in (3.1.4) and applying (3.1.6) yields
\begin{align}
    &(\sigma_\mu{}^{\nu \rho})_a{}^{\dot b} \partial_\nu \psi_{\rho \dot b} = [(\sigma_\mu \tilde \sigma^{\nu \rho})_a{}^{\dot b} - (\delta_\mu^{[\nu} \sigma^{\rho]})_a{}^{\dot b}]  \partial_\nu \psi_{\rho \dot b}\tag*{}\\
    &\qquad \qquad \quad \:\:\ = (\sigma_\mu)_{a}{}^{\dot c}{ \cancel{(\tilde \sigma^{\nu \rho})_{\dot c}{}^{\dot b} \partial_\nu \psi_{\rho \dot b}}} \quad - \:\:\delta_\mu^{[\nu}( \sigma^{\rho]})_a{}^{\dot b}  \partial_\nu \psi_{\rho \dot b} \tag*{}\\
    \implies \quad   &\delta_\mu^{[\nu} (\sigma^{\rho]})_a{}^{\dot b}  \partial_\nu \psi_{\rho \dot b} = 0
\end{align}
Under the standard $\sigma$-trace gauge condition $\sigma^\rho \psi_\rho = 0$, this relation reduces to the Dirac-like form
\begin{align}
    \delta_\mu^{[\nu} (\sigma^{\rho]})_a{}^{\dot b}  \partial_\nu \psi_{\rho \dot b} &= \underbrace{\cancel{\partial_\mu\bigl[(\sigma^\rho)_a{}^{\dot b} \psi_{\rho \dot b}\bigr]}}_{\text{Gauge Term}} - (\sigma^\nu)_a{}^{\dot b}\partial_\nu \psi_{\mu \dot b}\tag*{}\\
    \implies \quad (\sigma^\nu)_a{}^{\dot b}\partial_\nu \psi_{\mu \dot b} &= 0
\end{align}
More generally, for any tensor  $X^{U\mu}_L$ with arbitrary index structure $U, L$, antisymmetrization must be preserved:
\begin{equation}X^{U\mu}_L\delta_\mu^{[\nu} (\sigma^{\rho]})_a{}^{\dot b}  \partial_\nu \psi_{\rho \dot b} =X^{U[\nu}_L (\sigma^{\rho]})_a{}^{\dot b}  \partial_\nu \psi_{\rho \dot b} = 0 \end{equation}
The Rarita-Schwinger equation (3.1.4) therefore encodes both the dynamical content of a massless spin-$3/2$ field and the redundancies associated with linearized local supersymmetry. The contracted relation (3.1.6) may be viewed as the $\tilde \sigma$-trace of the equation of motion, while the decomposition isolates the $ \sigma$-trace contribution, making explicit how the Dirac-like form (3.1.8) emerges.

\subsection{Field Transformations}
The most general set of transformations consistent with gauge invariance, engineering dimension,
and reality conditions is

\begin{align}
     Q_a h_{\mu \nu} &= a_1(\sigma_{(\mu})_a{}^{\dot b} \psi_{\nu) \dot b} \\
      Q_a B_{\mu \nu} &= a_2 (\sigma_{[\mu })_a{}^{\dot b}\psi_{\nu] \dot b} + c_2(\sigma_{\mu \nu})_a{}^{b} \chi_b\\
       Q_a \phi &= c_1\chi_a \\
      Q_a \chi_b &=  if_0 (\sigma^{\rho})_{ba} \partial_{\rho} \phi + ie_0 (\sigma^{[3]})_{b a} A_{[3]}\\
      Q_a \psi_{\mu \dot{b}} &= id_0(\tilde \sigma^{\nu \rho})_{\dot{b} a}\partial_\nu h_{\rho
     \mu} + ie_1 ( \tilde \sigma^{\nu \rho })_{\dot b a} \partial_\nu B_{\rho\mu} + ie_2(\tilde \sigma_\mu\sigma^{[3]})_{\dot b a}A_{[3]}
\end{align}
Here $Q_a$ denotes the linearized supersymmetry generator acting on component fields, so that a
variation with constant Majorana--Weyl parameter $\epsilon^a$ is
$\delta_\epsilon = \epsilon^a Q_a$. All transformation laws are given to leading order in fermions
and with at most one derivative, as appropriate to the quadratic two-derivative free-field action.
The coefficients $a_1, a_2, c_1, c_2, f_0, e_0, d_0, e_1, e_2$ are real and a priori undetermined.

The allowed $\sigma$-matrix structures are constrained by ten-dimensional chirality, mass dimension, and Lorentz covariance. In particular, the dilatino variation admits only the minimal one- and three-index $\sigma$-matrix couplings at this order, while the gravitino variation may include $\tilde\sigma^{\nu\rho}$ contracted with derivatives of the two-form or, equivalently, with the gauge-invariant three-form field strength.

We may fix the overall normalization of the fermions by setting $a_1=c_2=1$, consistent with the
canonical kinetic terms of Sec.~3.1. This choice does not affect the physical content of the
linearized theory. Requiring on-shell closure of the supersymmetry algebra on all fields together with the
normalization of the supercurrent fixes the remaining coefficients uniquely:
\begin{equation}
\begin{alignedat}{3}
a_1&=1            &\qquad c_1&=\sqrt{8}            &\qquad d_0&=-\tfrac12\\
a_2&=1            &       f_0&=\tfrac{1}{\sqrt{8}}  &       e_1&=-\tfrac12\\
c_2&=1            &       e_0&=-\tfrac{1}{8}       &       e_2&=\tfrac{1}{16}
\end{alignedat}
\end{equation}
These coefficients determine the final form of the transformation laws in
Eqs.~(3.2.1)--(3.2.5):
\begin{align}
     Q_a h_{\mu \nu} &= (\sigma_{(\mu})_a{}^{\dot b} \psi_{\nu) \dot b} \\
      Q_a B_{\mu \nu} &=  (\sigma_{[\mu })_a{}^{\dot b}\psi_{\nu] \dot b} + (\sigma_{\mu \nu})_a{}^{b} \chi_b\\
       Q_a \phi &=  \sqrt{8} \chi_a \\
      Q_a \chi_b &=  \frac{i}{\sqrt{8}}(\sigma^{\rho})_{ba} \partial_{\rho} \phi -\frac{i}{8} (\sigma^{[3]})_{b a} A_{[3]}\\
      Q_a \psi_{\mu \dot{b}} &= -\frac{i}{2}(\tilde \sigma^{\nu \rho})_{\dot{b} a}\partial_\nu h_{\rho
     \mu} -\frac{i}{2} ( \tilde \sigma^{\nu \rho })_{\dot b a} \partial_\nu B_{\rho\mu} + \frac{i}{16}(\tilde \sigma_\mu\sigma^{[3]})_{\dot b a}A_{[3]}
\end{align}
A schematic discussion of the resulting algebra and its closure conditions is presented in the
next subsection.

\subsection{Closure of the Superalgebra}
We require on-shell closure of the linearized supersymmetry algebra on each field. In particular, for any component field $\Phi$,
\begin{equation}\{Q_a, Q_c\} \Phi\quad =\quad  2i\cancel \partial_{ac}\Phi \quad+\quad \delta^{\rm(gauge)}_{ac}\Phi\quad+\quad \delta^{\rm(EOM)}_{ac}\Phi \end{equation}
where $\cancel \partial_{ac}\equiv (\sigma^\mu)_{ac}\partial_\mu$ generates translations and
$\delta^{\rm(gauge)}_{ac}$ denotes the appropriate linearized gauge transformations of
$h_{\mu\nu}$ and $B_{\mu\nu}$. The fermionic contributions $\delta^{\rm(EOM)}_{ac}$ vanish
upon imposing the free equations of motion derived in Sec.~3.1. We present the essential results
here; representative intermediate steps,
$\sigma$-matrix manipulations and the coefficient constraints implied by closure are collected in Appendix~B.

At the level of generators this realizes the familiar schematic structure
$\{Q,Q\}\sim \Gamma^\mu P_\mu$, supplemented by field-dependent gauge transformations
compatible with the linearized symmetries of $h_{\mu\nu}$ and $B_{\mu\nu}$.

The gauge terms appearing below are organized as total derivatives and are to be interpreted as
linearized diffeomorphisms for $h_{\mu\nu}$ and Abelian two-form gauge transformations for $B_{\mu\nu}$. The appearance of these total-derivative terms is the expected manifestation of
field-dependent gauge parameters induced by two successive supersymmetry transformations,
and provides a check that the commutator respects the linearized gauge structure of the theory.

\paragraph{Bosonic sector.}
We first verify closure on $(h_{\mu\nu},B_{\mu\nu},\phi)$.
The resulting commutators exhibit translations plus the expected linearized gauge transformations. Intermediate steps and coefficient constraints can be found in Appendix~B.
\begin{align}
    &\{Q_a, Q_c\} h_{\mu \nu} \: = 2i\cancel{\partial}_{ac}h_{\mu\nu}+ \partial_{{(\mu|}}\Bigl[-i(\sigma^{\rho})_{ac}h_{\rho|{\nu)}}-i(\sigma^{\rho})_{ac}B_{\rho |{\nu)}}\Bigr]\\
    &\{Q_a, Q_c\} B_{\mu \nu}
     = 2i\cancel \partial_{ac}B_{\mu \nu}+ \partial_{{[\mu|}}\left[-i(\sigma^{\rho})_{ac}B_{\rho |{\nu]}} + 2i(\sigma^\rho)_{ac}h_{\rho |{\nu]}} - \frac{i}{\sqrt 2}(\sigma_{|{\nu]}})_{ac}\phi\right]\\
    &\{Q_a, Q_c\} \phi
    \ \: \: \: =  2i \cancel \partial_{ac}\phi
\end{align}

\paragraph{Fermionic sector.}
The corresponding computations for $\psi_{\mu\dot b}$ and $\chi_b$ require
use of the ten-dimensional Fierz identity (B.11). The structure of this identity has significant effects on the resulting non-closure geometry of the fermionic fields.

Recall $A_{\mu\nu\rho}=\frac{1}{3!}\partial_{[\mu}B_{\nu\rho]}$.
For any tensor $X^{U\mu\nu\rho}_L$ totally antisymmetric in $\mu\nu\rho$,
\begin{align}
X^{U\mu\nu\rho}_L A_{\mu\nu\rho}
= X^{U\mu\nu\rho}_L \partial_\mu B_{\nu\rho}.
\end{align}
This identity is used below to streamline several expressions.

\begin{align*}
    \{Q_a, Q_c\}\psi_{\mu \dot b}
    &= 2i\cancel \partial_{ac}\psi_{\mu \dot b} - \frac{7i}{8}(\sigma_{[1]})_{ac}(\tilde \sigma^{[1]} \sigma^\nu)_{\dot b}{}^{\dot d}\partial_\nu \psi_{\mu \dot d}+\frac{i}{120}(\sigma_{[5]})_{ac}(\tilde \sigma^{[5]} \sigma^\nu)_{\dot b}{}^{\dot d}\partial_\nu \psi_{\mu \dot d}\tag{3.3.6}\\
    &\quad\tag*{} -\frac{i}{2}\biggl[\frac{1}{8}(\sigma_{[1]})_{ac}(\tilde\sigma_\mu \sigma^{[1]} \tilde \sigma^\nu)_{\dot b}{}^d - \frac{7}{4}(\sigma_\mu)_{ac}(\tilde \sigma^\nu)_{\dot b}{}^d \\
    &\qquad \qquad\tag*{}- \frac{1}{16 \times 5!}(\sigma_{[5]})_{ac}(\tilde \sigma_\mu\sigma^{[5]}\tilde \sigma^\nu)_{\dot b}{}^d+\frac{1}{8 \times 4!}(\sigma_{[4]\mu})_{ac} (\tilde \sigma^{[4]}\tilde \sigma^\nu)_{\dot b}{}^d \biggr]\partial_\nu \chi_d\\
    &\quad\tag*{} + \frac{i}{16}\biggl[-(\sigma_{[1]})_{ac}(\tilde \sigma_{\mu}\sigma^{{[\nu|}}\tilde \sigma^{[1]}\sigma^{{|\delta]}})_{\dot b}{}^{\dot d}- 2(\sigma^{{[\nu|}})_{ac}(\tilde \sigma_\mu \sigma^{{|\delta]}})_{\dot b}{}^{\dot d}\\
    &\qquad\qquad\tag*{} +\frac{1}{2}(\sigma_{[1]})_{ac}(\tilde \sigma_\mu \sigma^{\nu [1]\delta})_{\dot b}{}^{\dot d}+\frac{1}{8 \times 5!}(\sigma_{[5]})_{ac}(\tilde \sigma_\mu  \sigma^{{[\nu|}} \tilde \sigma^{[5]} \sigma^{{|\delta]}})_{\dot b}{}^{\dot d}
    \biggr]\partial_\nu \psi_{\delta \dot d}\\
    &\quad\tag*{} -\frac{i}{2} \left[\frac{13}{8}(\sigma_{[1]})_{ac}(\tilde \sigma^{[1]})_{\dot b}{}^d  - \frac{11}{16 \times 5!}(\sigma_{[5]})_{ac}(\tilde \sigma^{[5]})_{\dot b}{}^d\right] \partial_\mu \chi_d\\
     &\quad\tag*{} +\partial_\mu \left[\frac{7i}{8}(\sigma_{[1]})_{ac}(\tilde \sigma^{[1]} \sigma^\nu)_{\dot b}{}^{\dot d}-\frac{i}{120}(\sigma_{[5]})_{ac}(\tilde \sigma^{[5]} \sigma^\nu)_{\dot b}{}^{\dot d} \right]   \psi_{\nu \dot d} \\
    \{Q_a, Q_c\} \chi_b
    &= 2i\cancel \partial_{ac}\chi_b -\frac{3i}{4}(\sigma_{[1]})_{ac}(\sigma^{[1]}\tilde \sigma^\nu)_{b}{}^d\partial_\nu \chi_d  -\frac{i}{8} \biggl[\frac{3}{2}(\sigma_{[1]})_{ac}(\sigma^{\nu \delta [1]})_b{}^{\dot d}\tag{3.3.7}\\
    &\quad+2(\sigma^{{[\delta|}})_{ac}(\sigma^{{|\nu]}})_b{}^{\dot d}+ \frac{1}{8 \times 5!}(\sigma_{[5]})_{ac}(\sigma^{{[\nu|}}\tilde \sigma^{[5]}\sigma^{{|\delta]}})_{b}{}^{\dot d}\biggr]\partial_\nu \psi_{\delta \dot d}
    \end{align*}
The resulting fermionic closures decompose naturally into the one- and five-index
Clifford channels, reflecting the standard ten-dimensional Fierz structure for
chiral spinors; the corresponding terms are proportional to the Rarita--Schwinger
and dilatino equations of motion and therefore vanish on shell.

\section{Supercurrent}
The supercurrent is obtained via the Noether procedure for global supersymmetry. For a constant Majorana--Weyl parameter $\epsilon^a$,
\begin{equation}
\partial_\mu\!\left(\epsilon^a (J^\mu)_a\right) = \partial_\mu\left[\sum_i \frac{\partial \mathcal L}{\partial (\partial_\mu \Phi_i)}\epsilon^a Q_a \Phi_i -\epsilon^a Q_a \mathcal L\right].
\end{equation}
The Noether supercurrent is defined up to terms of the form $(J^\nu)_a \to (J^\nu)_a + \partial_\mu U^{[\mu\nu]}{}_a$ which do not modify the conserved supercharge. We adopt the representative from this equivalence class that is convenient for matching the linearized algebra. Requiring that the supercurrent obtained from this variation be compatible with the closure relations of Sec.~3.3 and with canonical normalization of the kinetic terms provides the remaining conditions needed to uniquely fix the relative normalizations of the transformation laws and the free Lagrangian.

Intermediate steps and coefficient constraints can be found in Appendix~C. A direct computation yields
\begin{align}
\begin{split}
    (J^\mu)_a &=
     -3 A^{\nu \mu \rho}(\sigma_{\mu \rho})_a{}^b \chi_b - 2\sqrt 2 ( \sigma^{ \nu\mu})_a{}^b \phi \partial_\mu \chi_b \\
    &\quad + 3\psi_\lambda{}^c (\sigma^\nu)_{ca} \partial^\lambda h -  2\psi_\mu{}^c (\sigma^\mu)_{ca} \partial^\nu h +  2\psi_\mu{}^c (\sigma^\mu)_{ca} \partial^\rho h^\nu{}_\rho \\
    &\quad +2 \psi_\mu{}^c (\sigma^\rho)_{ca} \partial^\nu h^\mu{}_\rho -  2\psi_\mu{}^c (\sigma^\nu)_{ca} \partial^\rho h^\mu{}_\rho\\
    &\quad -(\sigma_{\mu})_{ca} \partial^\mu  h \psi^{\nu c} -2(\sigma^\mu)_{ca} \partial^\rho h_{\rho \mu}\psi^{\nu c}\\
    &\quad+2\psi_{\mu}{}^c(\sigma^{\mu \rho \lambda})_{ca}\partial_\lambda B^{\nu}{}_{ \rho}-2\psi^{\mu c}(\sigma^{ \nu \rho \lambda})_{ca}\partial_\lambda B_{\mu \rho}\\
    &\quad + \psi_{\mu}{}^c(\sigma^{\nu \rho \delta})_{ca}\partial^\mu B_{\delta \rho}-\psi_{\mu}{}^c(\sigma^{\mu \rho \delta })_{ca}\partial^\nu B_{\delta \rho}\\
    &\quad +\psi_\mu{}^c (\sigma^{\mu})_{ca}   \partial_\rho B^{\nu \rho} - 3\psi_{\mu}{}^c (\sigma^{[\mu }{}_{[2]})_{ca}A^{\nu][2]} \\
    &\quad-2\partial^\nu h^{\mu \rho} \left(\sigma_{\mu}\right)_a{}^{\dot b}\psi_{\rho \dot b} +2\partial^\nu h \left(\sigma^{\mu}\right)_a{}^{\dot b}\psi_{\mu \dot b}-2\partial_\beta h^{\nu \beta} \left(\sigma^{\mu}\right)_a{}^{\dot b}\psi_{\mu \dot b}\\&\quad-\partial_{\mu} h (\sigma^{(\nu})_a{}^{\dot b}\psi^{\mu)}{}_{ \dot b}+2\partial^\alpha h_{\alpha \mu}(\sigma^{(\nu})_a{}^{\dot b}\psi^{\mu)}{}_{\dot b}
\end{split}
\end{align}
Using the free equations of motion, the resulting supercurrent is conserved on shell,
\begin{equation}
\partial_\nu (J^\nu)_a = 0,
\end{equation}
consistent with global supersymmetry of the free action.

\section{Non-Closure Terms}
We collect here the fully determined linearized transformation laws and display the resulting supersymmetry algebra in a form that makes the gauge and on-shell structure explicit. The supersymmetry transformations with determined coefficients are relisted here for ease of reference. In the gravitino transformation we have used a standard $\sigma$-matrix identity in Eq.~(5.0.5) to present the result in a form suited for comparison with \cite{GatesNishino:1986a}.
\begin{align}
     Q_a h_{\mu \nu} &= (\sigma_{(\mu})_a{}^{\dot b} \psi_{\nu) \dot b} \\
      Q_a B_{\mu \nu} &=  (\sigma_{[\mu })_a{}^{\dot b}\psi_{\nu] \dot b} + (\sigma_{\mu \nu})_a{}^{b} \chi_b\\
       Q_a \phi &=  \sqrt{8} \chi_a \\
      Q_a \chi_b &=  \frac{i}{\sqrt{8}}(\sigma^{\rho})_{ba} \partial_{\rho} \phi -\frac{i}{8} (\sigma^{[3]})_{b a} A_{[3]}\\
      Q_a \psi_{\mu \dot{b}} &= -\frac{i}{2}(\tilde \sigma^{\nu \rho})_{\dot{b} a}\partial_\nu h_{\rho
     \mu} -\frac{i}{8} ( \tilde \sigma^{[3] }\sigma_\mu)_{\dot b a} A_{[3]} - \frac{i}{16}(\tilde \sigma_\mu\sigma^{[3]})_{\dot b a}A_{[3]} - \frac{i}{4} (\sigma^{\rho \eta})_{\dot b a} \partial_\mu B_{\rho \eta}
\end{align}
The superalgebra closes as
\begin{align}
    \{Q_a, Q_c\} h_{\mu \nu} &= 2i\cancel{\partial}_{ac}h_{\mu\nu} - i \partial_{{(\mu}} \Xi_{{\nu)} a c}\\
    \{Q_a, Q_c\} B_{\mu \nu} &= 2i\cancel{\partial}_{ac}B_{\mu\nu} - i \partial_{{[\mu}} \Lambda_{{\nu]} a c} \\
     \{Q_c, Q_a\} \phi &= 2i\cancel \partial_{ac} \phi \\
     \{Q_c, Q_a\} \psi_{\mu \dot b} &= 2i\cancel \partial_{ac} \psi_{\mu \dot b} + i \mathcal{\tilde{E}}_{\mu a c \dot b} + i\partial_\mu \Omega_{a c \dot b} \\
    \{Q_a, Q_c\} \chi_b &= 2i\cancel \partial_{ac} \chi_b + i\zeta_{a c b}
\end{align}
The gauge contributions in the algebra act on the fields as
\begin{align}
\delta^{\rm(gauge)}_{ac} h_{\mu\nu} &= -i\,\partial_{(\mu}\Xi_{\nu)ac},\\
\delta^{\rm(gauge)}_{ac} B_{\mu\nu} &= -i\,\partial_{[\mu}\Lambda_{\nu]ac},\\
\delta^{\rm(gauge)}_{ac} \psi_{\mu\dot b} &= i\partial_\mu \Omega_{ac\dot b},
\end{align}
with field-dependent parameters $\Xi$, $\Lambda$, and $\Omega$ given by
\begin{align}
   \Xi_{\nu a c} &= (\sigma^{\rho})_{ac}h_{\rho \nu} + (\sigma^{\rho})_{ac}B_{\rho \nu} \\
    \Lambda_{\nu a c} &= -(\sigma^{\rho})_{ac}B_{\rho \nu} + 2(\sigma^\rho)_{ac}h_{\rho \nu} - \frac{2}{\sqrt{8}}(\sigma_{\nu})_{ac}\phi \\
    \begin{split}
   \Omega_{a c \dot b} &=  \left[\frac{13}{8}(\sigma_{[1]})_{ac}(\tilde \sigma^{[1]})_{\dot b}{}^d  - \frac{11}{16 \times 5!}(\sigma_{[5]})_{ac}(\tilde \sigma^{[5]})_{\dot b}{}^d\right]  \chi_d\\
    &\quad +\left[\frac{7}{8}(\sigma_{[1]})_{ac}(\tilde \sigma^{[1]} \sigma^\nu)_{\dot b}{}^{\dot d}-\frac{1}{120}(\sigma_{[5]})_{ac}(\tilde \sigma^{[5]} \sigma^\nu)_{\dot b}{}^{\dot d} \right]   \psi_{\nu \dot d}
    \end{split}
\end{align}
The remaining non-closure contributions can be organized into tensors proportional to the
Rarita--Schwinger and dilatino equations of motion.
That the bosonic commutators close without equation-of-motion terms, while the fermionic
commutators require them, is the expected feature of ten-dimensional $\mathcal N=1$ supergravity
at the two-derivative level, where closure is generically on shell in the absence of auxiliary
fields. The non-closure contributions act on the fields as
\begin{align}
\delta^{\rm(EOM)}_{ac} \psi_{\mu\dot b} &= i\mathcal{ \tilde{E}}_{\mu a c \dot b},\\
\delta^{\rm(EOM)}_{ac} \chi_{\mu b} &= i\zeta_{acb},
\end{align}
with field-dependent parameters $\mathcal{ \tilde{E}}$ and $\zeta$ given by
\begin{align*}
    \mathcal{\tilde{E}}_{\mu a c \dot b } &= \tag{5.0.19}- \frac{7}{8}(\sigma_{[1]})_{ac}(\tilde \sigma^{[1]} \sigma^\nu)_{\dot b}{}^{\dot d}\partial_\nu \psi_{\mu \dot d}+\frac{1}{120}(\sigma_{[5]})_{ac}(\tilde \sigma^{[5]} \sigma^\nu)_{\dot b}{}^{\dot d}\partial_\nu \psi_{\mu \dot d}\\
    &\quad -\frac{1}{2}\biggl[\frac{1}{8}(\sigma_{[1]})_{ac}(\tilde\sigma_\mu \sigma^{[1]} \tilde \sigma^\nu)_{\dot b}{}^d - \frac{7}{4}(\sigma_\mu)_{ac}(\tilde \sigma^\nu)_{\dot b}{}^d \\
    &\qquad \qquad- \frac{1}{16 \times 5!}(\sigma_{[5]})_{ac}(\tilde \sigma_\mu\sigma^{[5]}\tilde \sigma^\nu)_{\dot b}{}^d+\frac{1}{8 \times 4!}(\sigma_{[4]\mu})_{ac} (\tilde \sigma^{[4]}\tilde \sigma^\nu)_{\dot b}{}^d \biggr]\partial_\nu \chi_d\\
    &\quad + \frac{1}{16}\biggl[-(\sigma_{[1]})_{ac}(\tilde \sigma_{\mu}\sigma^{{[\nu|}}\tilde \sigma^{[1]}\sigma^{{|\delta]}})_{\dot b}{}^{\dot d}- 2(\sigma^{{[\nu|}})_{ac}(\tilde \sigma_\mu \sigma^{{|\delta]}})_{\dot b}{}^{\dot d}\\
    &\qquad\qquad +\frac{1}{2}(\sigma_{[1]})_{ac}(\tilde \sigma_\mu \sigma^{\nu [1]\delta})_{\dot b}{}^{\dot d}+\frac{1}{8 \times 5!}(\sigma_{[5]})_{ac}(\tilde \sigma_\mu  \sigma^{{[\nu|}} \tilde \sigma^{[5]} \sigma^{{|\delta]}})_{\dot b}{}^{\dot d}
    \biggr]\partial_\nu \psi_{\delta \dot d}\\\\
     \zeta_{a c b} &= -\frac{3}{4}(\sigma_{[1]})_{ac}(\sigma^{[1]}\tilde \sigma^\nu)_{b}{}^d\partial_\nu \chi_d  -  \frac{1}{8}\biggl[\frac{3}{2}(\sigma_{[1]})_{ac}(\sigma^{\nu \delta [1]})_b{}^{\dot d}\tag{5.0.20}\\
    &\quad+2(\sigma^{{[\delta|}})_{ac}(\sigma^{{|\nu]}})_b{}^{\dot d}+ \frac{1}{8 \times 5!}(\sigma_{[5]})_{ac}(\sigma^{{[\nu|}}\tilde \sigma^{[5]}\sigma^{{|\delta]}})_{b}{}^{\dot d}\biggr]\partial_\nu \psi_{\delta \dot d}
\end{align*}
\section{L \& R (Adjacency) Matrices}
We now have all the information needed to extract our right and left adjacency matrices, $R_I$ and $L_I$. After 0-brane reduction and the gauge choices specified below, we expect the $R_I$ to be $n_b \times n_f$ and the $L_I$ to be $n_f \times n_b$ dimensional. Here $n_b$ and $n_f$ denote the numbers of independent component fields retained after the reductions specified below, rather than the standard ten-dimensional propagating degrees of freedom. By projecting onto the 0-brane during construction of the $R_I$ and $L_I$ we remove spatial dependency of our fields. Explicitly, under gauge transformation $h_{0 \nu}$ transforms like
\begin{equation}\delta_G h_{0 \nu} = \partial_{(0} \lambda_{\nu)} = \partial_0 \lambda_\nu\end{equation}
If $h_{0 \nu} \not = 0$ we can always make the gauge transformation
\begin{equation}h_{0 \nu} \to h_{0\nu} + \partial_0 \lambda_\nu\qquad \text{where} \qquad \lambda_\nu = - \int dt h_{0 \nu}\end{equation}
in order to set $h_{0\nu} = 0$ (10 components), and similarly for $B_{0 \nu}$ (9 components), assuming boundary conditions such that the time integral is well defined. This choice of gauge reduces $n_b$ by 19 and gives
\begin{align}n_b &= \text{dof}(h_{\mu \nu}) + \text{dof}(B_{\mu \nu}) + \text{dof}(\phi)=55 + 45 + 1 - 19 = 82
\end{align}
The fermions simply satisfy
\begin{align}
    n_f = \text{dof}(\psi_{\mu \dot a})+\text{dof}(\chi_a) = 160 +16 = 176
\end{align}
Thus we expect $82 \times 176$ and $176 \times 82$ dimensional matrices.

\indent With our counting done and our supermultiplet explicitly completed we are now poised to find $R_I$ and $L_I$ via the process given by \cite{Chappell_2013}.
Projection onto the 0-brane corresponds to dimensional reduction to one time dimension, so that all fields depend only on $t$ and $\partial_\mu\to\delta_\mu^0\,\partial_0$. We order the independent components of $h_{\mu\nu}$ with $\mu\le\nu$ and of $B_{\mu\nu}$ with $\mu<\nu$. We recast our equations in component form for each supercharge $Q_a$. E.g.,
\begin{equation}
\begin{array}{lll}
\begin{aligned}
Q_1 h_{11} &= 2 \psi_{1(6)} \\
&\: \ \vdots \\
Q_1 h_{47} &= - \psi_{4(4)} - \psi_{7(5)} \\
&\: \ \vdots \\
Q_1 h_{99} &= 2 \psi_{9(11)}
\end{aligned}
&
\qquad
\begin{aligned}
Q_1 B_{12} &= -\psi_{1(8)} + \psi_{2(6)} + \chi_9 \\
&\:\ \vdots \\
Q_1 B_{37} &= \psi_{3 ( 4)} - \psi_{7(7)} +  \chi_{16}\\
&\:\ \vdots \\
Q_1 B_{89} &= \psi_{8(11)} + \psi_{9(3)} + \chi_3
\end{aligned}
&
\qquad
\begin{aligned}
Q_1 \phi &= \sqrt{8}\chi_1
\end{aligned}
\end{array}
\end{equation}
where we have bracketed the spinor index of the gravitino for ease of notation. Similarly for the fermionic sector,
\begin{equation}
\begin{array}{ll}
\begin{aligned}
Q_1 \psi_{1(1)} &= \frac{i}{2}\partial_0 h_{1 9} + \frac{i}{16} \partial_0 B_{1 \, 9} \\
&\: \ \vdots \\
Q_1 \psi_{4(16)} &= \frac{i}{2}\partial_0 h_{1 \, 4} - \frac{i}{16} \partial_0 B_{1 4} + \frac{3i}{16}  \partial_0[ B_{2 3} + B_{5 6} + B_{7 8}] \\
&\: \ \vdots \\
Q_1 \psi_{9 (16)} &= \frac{i}{2} \partial_0 h_{1 9} - \frac{i}{16} \partial_0 B_{1 9}
\end{aligned}
&
\qquad
\begin{aligned}
Q_1 \chi_1 &= \frac{i}{\sqrt{8}} \partial_0 \phi  \\
&\: \ \vdots \\
Q_1 \chi_9 &= -\frac{i}{8} \partial_0 B_{8 9} \\
&\: \ \vdots \\
Q_1 \chi_{16} &= -\frac{i}{8} \partial_0 B_{1 9}
\end{aligned}
\end{array}
\end{equation}
and so on for $Q_2, \ldots, Q_{16}$. We compile all bosons and fermions into two vectors:
\begin{equation}\Phi = (h_{11}, \ldots, h_{99}, B_{12}, \ldots, B_{89}, \phi), \qquad \Psi = \frac{1}{i}(\psi_{0(1)}, \ldots, \psi_{9(16)}, \chi_1, \ldots, \chi_{16})\end{equation}
where $\Phi$ is taken to consist only of the gauge-fixed 0-brane bosons,
i.e.\ the components with purely spatial indices $i,j=1,\ldots,9$. With these conventions we may write
\begin{equation}
(Q_{ I} \Phi)_j = i (L_{ I})_j{}^{\hat k}\,\Psi_{\hat k},
\qquad
(Q_{ I} \Psi)_{\hat j} = (R_{ I})_{\hat j}{}^{k}\,\dot{\Phi}_{k},
\end{equation}
where the index $I$ runs from 1 to 16. The factor of $i$ in the first equation guarantees that the adjacency matrices are real in this basis. They satisfy the 0-brane (1 dimensional) form of the Garden algebra:
\begin{equation}
L_{ \bm I} R_{ \bm J} + L_{ \bm J} R_{ \bm I} = 2 \delta_{\bm I  \bm J}\mathbb I_{82 \times 82}
\end{equation}
Explicit representations of these matrices can be found on \href{https://github.com/mcmulaz/Super-Sym}{github}\footnote{https://github.com/mcmulaz/Super-Sym}. The counterpart anti-commutator equation satisfies
\begin{equation}
R_{\bm I} L_{\bm J} + R_{\bm J}L_{\bm I} = 2\delta_{\bm I \bm J}\mathbb{I}_{176 \times 176} + 2 \mathcal{E}_{\bm I \bm J} \mathbb{I}_{176 \times 176}
\end{equation}
And finally, we also present the associated commutator equations
\begin{equation}
     L_{\bm I} R_{\bm J} - L_{\bm J}R_{\bm I} = 2\mathcal{B}_{\bm I \bm J}^{ (B)}
\end{equation}
\begin{equation}
    R_{\bm I} L_{\bm J} - R_{\bm J}L_{\bm I} = 2\mathcal{B}_{\bm I \bm J}^{ (F)}
\end{equation}
$2\mathcal{B}_{\bm I \bm J}^{ (B)}$
and $2\mathcal{B}_{\bm I \bm J}^{ (F)}$ are the associated bosonic and fermionic holoraumy tensors respectively (see \cite{gates2024moderntensorspinorsymbolicalgebra},\cite{holo1}, \cite{holo2}).
The non-closure tensor $\mathcal{E}_{IJ}$ encodes the residual non-closure of the on-shell algebra, including remnant gauge transformations and fermionic equation-of-motion terms. An off-shell completion corresponds to enlarging the field content and/or modifying the transformation laws so as to eliminate $\mathcal{E}_{IJ}$.
We expect that making the explicit 0-brane algebra available will facilitate further progress on the off-shell problem for $10D$, $\mathcal N=1$ linearized supergravity.
\section{Conclusion and Future Pathways}

The main result of this work is to report the creation of a new `library' of the $L_{\bm I}$ and $R_{\bm J}$
matrices that for the first time have been directly derived from the {\em {on}}-{\em {shell}} description of the ten dimensional
on-shell supergravity supermultiplet.  The sixteen $L_{\bm I}$ matrices are 82 $\times$176 by size and the sixteen $R_{\bm I}$ matrices are 176 $\times$ 82 by size.  Rather than explicitly demonstrating them, their forms are accessible via the link \href{https://github.com/mcmulaz/Super-Sym}{github}.

Now that these matrices have been uncovered, the path forward involves additional forensic analysis in
a program that has identifiable steps.  We believe it is useful to describe these:

\subsection{Leveraging Ten Dimensional Supergravity Superspace Results}

In the work of \cite{GatesVashakidze:1987I,GatesNishino:1987II} the first superspace modification
of the lowest order open superstring correction that accommodated the Born-Infeld action was
obtained in the context of the 10D supersymmetric Yang-Mills theory. The usual on-shell component fields satisfy the equations
\begin{align}
Q_a v_{\mu}{}^{\Hat I} &= i (\sigma_{\mu})_{a \, b}  \lambda^{b}{}^{\Hat I} ~~~,  \\
Q_a \lambda^{b}{}^{\Hat I}  &=  \, (\sigma^{\mu \, \nu})_{a}{}^{\, b}  F_{\mu \, \nu }{}^{\Hat I} ~~~,
\end{align}
with a gauge field $v_{\mu}{}^{\Hat I}$ for a gauge group with generators $t_{\Hat I}$ with ${\Hat I} $
= $1, \, ..., n$ and a corresponding gaugino field $\lambda^{b}{}^{\Hat I} $.  However the addition of the
open-superstring correction demanded a new auxiliary bosonic field denoted by $ f_{[5]}{}^{{\Hat I}}$.  The
supergeometrical structure requires this to appear in the superspace fiber bundle geometry via the
structure (the following equation conventions are those in \cite{GatesVashakidze:1987I,GatesNishino:1987II})
\begin{equation}
\begin{aligned}
{\left[\nabla_{\alpha}, \nabla_{\beta}\right\} } & =i \sigma_{\alpha \beta}^{\underline{\varepsilon}} \nabla_{\underline{c}}+g \frac{1}{5!}\left(\sigma^{[5]}\right)_{\alpha \beta} f_{[5]}{}^{{\Hat I}} t_{\hat{I}} ~~,~~
{\left[\nabla_{\alpha}, \nabla_{\underline{b}}\right\} }  =-i g\left[\lambda_{\alpha \underline{b}}^{\hat{I}}+i \sqrt{\frac{1}{2}} \sigma_{\underline{b} \alpha \gamma} W^{\hat{I} \gamma}\right] t_{\hat{I}} ~~, \\
F_{\alpha \beta}{}^{\Hat I}  & =  -i \frac{1}{5!}\left(\sigma^{[5]}\right)_{\alpha \beta} f_{[5]}{}^{\Hat I} ~~,~~
{\left[\nabla_{\underline{a}}, \nabla_{\underline{b}}\right\} }  =i g F_{\underline{a} \underline{b}} {}^{\Hat I} t_{\Hat I} ~~~,~~~
 \nabla_{\underline{a}}=i \frac{1}{16} \sigma_{\underline{a}}^{\alpha \beta}\left[\nabla_{\alpha}, \nabla_{\beta}\right\}
~~,~~
\end{aligned}
\end{equation}
where the super Yang-Mills Field Strength super-tensor components $ f_{\underline{a}_{1} \ldots
\underline{a}_{5}}{}^{\Hat I}$ and $\lambda_{\alpha \underline{b}}{}^{\Hat I} $ must satisfy
\begin{equation}
\begin{aligned}
f_{\underline{a}_{1} \ldots \underline{a}_{5}}{}^{\Hat I} & =-\frac{1}{5!} \varepsilon_{\underline{a}_{1} \ldots \underline{a}_{s} \underline{b}_{1} \ldots \underline{b}_{5}} f^{b_{1} \ldots \underline{b}_{s} \hat{I}} ~~,~~
\sigma^{\underline{b} \alpha \beta} \lambda_{\alpha \underline{b}}{}^{\Hat I}   =0 ~~. \\
\end{aligned}
\end{equation}
The superspace Bianchi identities then require,
\begin{equation}
\begin{aligned}
\nabla_{\alpha} f_{[5]}{}^{\Hat I}= & \frac{1}{28}\left(\sigma_{[5]}\right)^{\beta \gamma}\left[\sigma_{\alpha \beta}^{\hat{c}} \lambda_{\gamma \underline{c}}^{\hat{I}}+\left(\sigma^{[3]}\right)_{\alpha \beta} \lambda_{\gamma[3]}{}^{\Hat I}\right]  ~~~,\\
\nabla_{\alpha} W{}^{{\Hat I} \beta}= & -\frac{1}{2} \sqrt{\frac{1}{2}}\left(\sigma^{b c}\right)_{\alpha}^{\beta}\left[F_{\underline{b c}}{}^{\Hat I}+d_{\underline{b c}}^{\hat{I}}\right]-\frac{1}{\sqrt{2} \cdot 4!}\left(\sigma^{[4]}\right)_{\alpha}^{\beta} d_{[4]}
{}^{\Hat I}
~~~, \\
\nabla_{\alpha} \lambda_{\beta \underline{c}}{}^{\Hat I}= & i\left(\sigma^{\underline{d}_{1} \underline{d}_{2} \underline{d}_{3}}\right)_{\alpha \beta} b_{\underline{c} \underline{d}_{1} \underline{d}_{2} \underline{d}_{3}}^{\hat{I}}+i \frac{1}{32}\left[5 \sigma_{\underline{c} \alpha \gamma}\left(\sigma^{[2]}\right)_{\beta}^{\gamma}+3\left(\sigma^{[2]}\right)_{\alpha}^{\gamma} \sigma_{\underline{c} \gamma \beta}\right] d_{[2]}{}^{\Hat I}  \\
& +i \frac{1}{8 \cdot 4!}\left[5 \sigma_{\underline{c} \alpha \gamma}\left(\sigma^{[4]}\right)_{\beta}^{\gamma}-\left(\sigma^{[4]}\right)_{\alpha}^{\gamma} \sigma_{\underline{c} \beta \gamma}\right] d_{[4]}^{\hat{I}} \\
& -i \frac{1}{4!}\left[\left(\sigma^{\underline{e}_{1} \underline{e}_{2} \underline{e}_{3}}\right)_{\alpha \beta} \nabla^{\underline{d}} f_{\underline{c d e_{1}} \underline{\underline{e}}_{2} \underline{e}_{3}}^{\hat{I}}+\frac{1}{10}\left(\sigma^{[5]}\right)_{\alpha \beta} \nabla_{\underline{c}} f_{[5]}^{\hat{I}]}\right]  ~~~, \\
\nabla_{\alpha} F_{\underline{b c}}{}^{\Hat I}= & -\nabla_{\left[\underline{b}^{\prime}\right.} \lambda_{\alpha \underline{c}]}^{\hat{I}}+i \sqrt{\frac{1}{2}} \sigma_{[\underline{b} \alpha \gamma} \nabla_{\underline{c}]} W^{\hat{I} \gamma}
~~~, \\
b_{\underline{d}_{1} \underline{d}_{2} \underline{d}_{3}}{}^{\Hat I}= & b_{\left[\underline{c d_{1}} \underline{d}_{2} \underline{d}_{3}\right]}{}^{\Hat I}=\sigma^{\underline{a} \alpha \beta} \lambda_{\beta \underline{a b c}}{}^{\Hat I}=0 ~~,
\end{aligned}
\end{equation}
these lead to spacetime equations of motion for the fermion,
\begin{equation}
\begin{gathered}
i \sigma_{\delta \beta}^{\underline{a}} \nabla_{\underline{a}} W^{\hat{I} \beta}=\frac{3}{35} \sqrt{\frac{1}{2}}\left(\sigma^{[2]}\right)_{\delta}^{\beta} \nabla_{\beta} d_{[2]}{}^{\Hat I}-\frac{1}{420} \sqrt{\frac{1}{2}}\left(\sigma^{[4]}\right)_{\delta}^{\beta} \nabla_{\beta} d_{[4]}{}^{\Hat I}+\frac{3}{35} \sqrt{\frac{1}{2}} \nabla^{\underline{a}} \lambda_{\delta \alpha}{}^{\Hat I} \\
i \sigma_{\delta \beta}^{\underline{a}} \nabla_{\underline{a}} W^{{\Hat I} \beta}=\frac{-1}{2 \cdot 7!\cdot \sqrt{2}}\left[i \sigma_{\underline{a}}^{\beta \varepsilon}\left(\sigma^{[4]}\right)_{\varepsilon}^{\gamma}\left(\sigma^{a b}\right)_{\delta}^{\lambda} \nabla_{\lambda} \nabla_{\beta} \nabla_{\gamma} f_{\underline{b}[4]}{}^{\Hat I}\right. \\
\left.+8\left(\sigma^{[4]}\right)_{\delta}^{\beta}\left\{\nabla^{\underline{b}}, \nabla_{\beta}\right\} f_{\underline{b}[4]}{}^{\Hat I}\right] ~~~, \\
\end{gathered}
\end{equation}
and spacetime equations of motion for the boson
\begin{equation}
\begin{aligned}
\nabla^{\underline{b}} F_{\underline{b} \underline{c}}{}^{\Hat I}= & i \frac{1}{2} g \sigma_{\underline{c} \alpha \beta} f_{\hat{J} \hat{K}}^{\hat{I}} W^{\hat{J} \alpha} W^{\hat{K} \beta}-\nabla^{\underline{b}} d_{\underline{b} \underline{c}}{}^{\hat{I}}-g \frac{1}{4} \sqrt{\frac{1}{2}} f_{\hat{J} \hat{K}}^{\hat{I}} W^{\hat{j} \beta} \lambda_{\beta \underline{c}}{}^{\hat{K}} \\
& +i \frac{3}{16} \cdot \frac{1}{35} \sigma_{\underline{c}}^{\alpha \gamma}\left(\sigma^{a b}\right)_{\gamma}^{\beta} \nabla_{\alpha} \nabla_{\beta} d_{\underline{a} \underline{b}}{}^{\Hat I}-i \frac{1}{16} \cdot \frac{1}{420} \sigma_{\underline{c}}^{\alpha \gamma}\left(\sigma^{\underline{b} 1} \cdots \underline{b}_{4}\right)_{\gamma}^{\beta} \nabla_{\alpha} \nabla_{\beta} d_{\underline{b}_{1} \ldots \underline{b}_{4}}{ }^{\hat{I}} \\
& +i \frac{3}{8} \cdot \frac{1}{35} \sigma_{\underline{c}}^{\alpha \beta} \nabla_{\alpha} \nabla^{\underline{c}} \lambda_{\beta \underline{c}}{}^{\Hat I} ~~~,  \\
\nabla^{\underline{b}} F_{\underline{b c}}{}^{\Hat I}= & i \frac{1}{2} g \sigma_{\underline{c} \alpha \beta} f^{\hat{I}}{ }_{\hat{J} \hat{K}} W^{\hat{J} \alpha} W^{\hat{K} \beta}+g \frac{1}{4 \cdot 5!\cdot \sqrt{2}} f^{\hat{I}}{ }_{\hat{J}} W^{\hat{J} \alpha}\left(\sigma^{[4]}\right)_{\alpha}{ }^{\beta} \nabla_{\beta} f_{\underline{c}[4]}{}^{\Hat I} \\
& +i \frac{1}{8 \cdot 5!} \sigma^{\underline{b} \alpha_{\gamma}}\left(\sigma^{[4]}\right)_{\gamma}{ }^{\beta} \nabla_{\underline{b}} \nabla_{\alpha} \nabla_{\beta} f_{\underline{c}[4]}{}^{\Hat I} \\
& +\frac{1}{32 \cdot 7!}\left[\sigma_{\underline{c}}^{\alpha \varepsilon}\left(\sigma^{a b}\right)_{\varepsilon}{ }^{\beta} \sigma_{\underline{a}}{ }^{\gamma \lambda}\left(\sigma^{[4]}\right)_{\lambda}{ }^{\delta} \nabla_{\alpha} \nabla_{\beta} \nabla_{\gamma} \nabla_{\delta} f_{\underline{b}[4]}{}^{\Hat I}\right. \\
& \left.-i 8 \sigma_{\underline{c}}^{\alpha \gamma}\left(\sigma^{[4]}\right)_{\gamma}{ }^{\beta} \nabla_{\alpha}\left\{\nabla^{\underline{b}}, \nabla_{\beta}\right\} f_{\underline{b}[4]}{}^{\Hat I}\right] ~~~.
\end{aligned}
\end{equation}

The final one of these equations delivers an astounding decree...if the spinor-spinor component $F_{\alpha
\beta}{}^{\Hat I}$ vanishes, the space-time component $F_{\underline a \underline b}{}^{\Hat I}$ satisfies the
spacetime equation of motion for a Yang-Mills field coupled to a current that is sourced by the gaugino $W^{\hat{I} \alpha}$.
The reason this is unexpected is that the condition $F_{\underline a \underline b}{}^{\Hat I}$ = 0 does {\em {not}} imply
space-time equations of motion in six and lower dimensions.

Now we can call upon the ``Breitenlohner observation'' \cite{Breitenlohner1977Geometric} where for the first time it was noted that introducing an `extra' spacetime vector index to a spin-one gauge field
in equations of a supermultiplet implies SUSY transformation laws in a supergravity supermultiplet.

Implementation of the  ``Breitenlohner observation'' here implies that the form of the spinor-spinor component
of the 10D superspace torsion description should be expected to take the form
\begin{equation}
T_{\alpha \beta}{}^{\underline c}   =  -i \frac{1}{5!}\left(\sigma^{[5]}\right)_{\alpha \beta} f_{[5]}{}^{\underline c}   ~~~,
\end{equation}
and thus a corresponding component field $  f_{[5]}{}^{\underline c} $ is implied for an off-shell 10D SG supermultiplet!

\subsection{Leveraging Ten Dimensional Adynkra Results}

In the work of \cite{Gates:2020Prepot10D}, a study was made directly on the possible supergravity
prepotentials that can lead to a superspace geometrical description that is of necessity off-shell.
This study was enabled by the use of a fairly new graphical constructs called adynkras.  The adinkras introduced in the 2000's \cite{FG:2005}, mathematically speaking, are the ``forgetful functors''' of the newer adynkras defined in the work of 2020's.  The latter carry the additional information about the higher dimensional
Lorentz group representations.

These results discussed in this chapter together with the sixteen $L_{\bm I}$ and $R_{\bm I}$ matrices uncovered in this work may be regarded as pieces of a large jig saw puzzle.  Conceptually, this ``jig saw puzzle'' can be explained by
looking back at a discussion of what adinkras uncovered about the 4D, $\cal N$ = 1 chiral supermultiplet shown in the work of \cite{Gates:2009G-1}.  That
work contains the diagram below.
\begin{figure}[h!]
    \centering
    \includegraphics[width=4.2in]{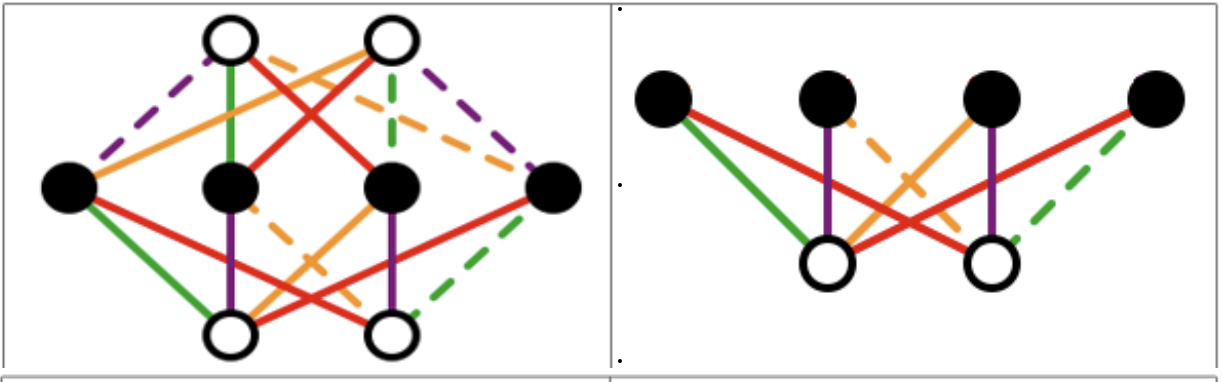}

    \caption{Off-shell and on-shell Adinkras for the 4D, $\mathcal{N}=1$ chiral supermultiplet.}
    \label{fig:adnkChi}
\end{figure}

To the left hand side of this diagram is the complete adinkra that describes the
off-shell chiral supermultiplet, while to the right hand side there is shown the on-shell adinkra version.  It is easy to see the on-shell version (on the right) `fits' as the lower half of the off-shell version (on the left) in the context of this simpler theory.  So diagrammatically it is easy to understand how the auxiliary fields solve the off-shell problem here.

The relevance of this discussion to the present work is that this paper has
derived the L-matrix and R-matrix data (the adjacency matrices data) for the
on-shell 10D, $\cal N$ = 1 SG theory.  This is the data equivalent to the rhs
image above.  The work in \cite{Gates:2020Prepot10D} contains the data
for several possible 10D, $\cal N$ = 1 SG prepotentials, accordingly these are
equivalent to the matrices that describe the leftmost image.  So the unsolved
problem is to derive how the on-shell data set (in this work) fits into which of the candidates SG prepotentials in \cite{Gates:2020Prepot10D}.  Unfortunately, solving this problem, at this time, is not possible due to an insufficiency of the appropriate codes.

A successful
merging of these pieces ought to lead to an off-shell superspace prepotential for ten dimensional
$\cal N$ = 1 supergravity.

\vspace{.05in}
\centering
\parbox{4in}{{\it ``
To deal with hyper-planes in a 14-dimensional space, visualize a 3-D space and say 'fourteen' to yourself very loudly. Everyone does it.'' \\ ${~}$
 ${~}$
\\ ${~}$ }\,\,-\,\, Geoffrey Hinton $~~~~~~~~~$}

\appendix
\section{Gamma Matrix Conventions}
\label{sec:gamma-matrices}
\numberwithin{equation}{section}
In this section we discuss the conventions used with our sigma matrices. We follow the conventions of \cite{cite-key}. \\
In $d$ dimensions the Clifford algebra is defined by a set of $d$ matrices $\gamma^\mu$ that satisfy
\begin{equation}\{\gamma^\mu, \gamma^\nu\} = 2\eta^{\mu \nu}\mathbb{I}\end{equation}
where the Minkowski metric is chosen to be mostly positive. In general these are $2^{d/2} \times 2^{d/2}$ complex matrices, but in 10D they satisfy Majorana-Weyl conditions. That is, they are real and can be written
\begin{equation}\gamma^\mu = \begin{pmatrix}
    0 & \sigma^\mu\\
    \tilde \sigma^\mu & 0
\end{pmatrix}\end{equation}
where the $\sigma^\mu$, $\tilde \sigma^\mu$ satisfy
\begin{equation}\sigma^\mu \tilde \sigma^\nu + \sigma^\nu \tilde \sigma^\mu = 2\eta^{\mu \nu} \mathbb I_{16 \times 16}, \qquad \tilde\sigma^\mu \sigma^\nu + \tilde\sigma^\nu  \sigma^\mu = 2\eta^{\mu \nu} \mathbb I_{16 \times 16}\end{equation}
This reduces our spinors to 16 components rather than 32. We represent spacetime indices with greek letters and spinor indices with latin letters. Left handed spinor indices are undotted and vice versa. To raise and lower spinor indices we use the spinor metric $C_{a \dot b}$
\begin{equation}
    \psi_{\mu}{}^a C_{a \dot b}= \psi_{\mu \dot b} \quad , \quad \psi_{\mu }{}^{\dot a} C_{b \dot a} = \psi_{\mu b}
\end{equation}
Its inverse $C^{a \dot b}$ is defined by
\begin{equation}
    C_{a \dot b}C^{c \dot b} = \delta_a^c  \quad, \quad C_{b \dot a}C^{b \dot c} = \delta_{\dot a}^{\dot c}
\end{equation}
Left handed sigma matrices are given by
\begin{equation}
    (\sigma^\mu)_{a b}, \qquad (\sigma^{\mu \nu \gamma})_{a b} = \frac{1}{3!}(\sigma^{[\mu}\tilde \sigma^{\nu}\sigma^{\gamma]})_{ab}, \qquad (\sigma^{\mu \nu \gamma \rho \delta})_{a b} = \frac{1}{5!}(\sigma^{[\mu}\tilde \sigma^\nu \sigma^\gamma \tilde \sigma^\rho \sigma^{\delta]})_{ab}
\end{equation}
The left handed mixed bispinors are defined similarly.
\begin{equation}
(\sigma^{\mu \nu })_{a \dot b}  = \frac{1}{2}(\sigma^{[\mu}\tilde \sigma^{\nu]})_{a \dot b}\quad , \quad (\sigma^{\mu \nu \gamma \rho})_{a \dot b} = \frac{1}{4!}(\sigma^{[\mu}\tilde \sigma^\nu \sigma^\gamma \tilde \sigma^{\rho]})_{a \dot b}
\end{equation}
Right handed sigma matrices and bispinors are denoted with a tilde, and are defined similarly to above with the role of left and right handed matrices interchanged.
\begin{equation}
(\tilde{\sigma}^\mu)_{\dot a \dot b} \quad , \quad (\tilde{\sigma}^{\mu \nu \gamma})_{\dot a \dot b} \quad , \quad (\tilde{\sigma}^{\mu \nu \gamma \rho \delta})_{\dot a \dot b}
\end{equation}
\begin{equation*}
(\tilde{\sigma}^{\mu \nu})_{ \dot a b} \quad , \quad (\tilde{\sigma}^{\mu \nu \gamma \rho})_{\dot a b}
\end{equation*}
Purely left handed matrices have the following index symmetries
\begin{align}
    (\sigma^\mu)_{a b} &=  (\sigma^\mu)_{b a} \\
    (\sigma^{\mu \nu \rho})_{a b} &=  -(\sigma^{\mu \nu \rho})_{b a} \\
    (\sigma^{\mu \nu \rho \gamma \delta})_{a b} &=  (\sigma^{\mu \nu \rho \gamma \delta})_{b a}
\end{align}
and similarly for the right handed matrices.
Equations A.9 - A.12, in addition to the Clifford algebra, yield an equivalent definition to those given in Appendix B of \cite{cite-key}. \\
The following Fierz identity is crucial to the closure calculation above, but it is the only such identity needed.
\begin{equation}
    \delta_{(a}{}^b\delta_{c)}{}^d = \frac{1}{8}\big{\{} (\sigma^{[1]})_{a c}(\tilde{\sigma}_{[1]})^{b d} + \frac{1}{2\times5!} (\sigma^{[5]})_{a c}(\tilde{\sigma}_{[5]})^{b d}
    \big{\}}
\end{equation}
The two following higher-index analogues of the Clifford equation (A.1) are also used at length.
\begin{align}
    (\sigma^\nu \tilde \sigma^{[4]})_{a b}(\sigma_{[4]})_{c \dot d} &= (\sigma^{[4]} \sigma^\nu )_{a b}(\sigma_{[4]})_{c \dot d} - 8(\sigma^{[3]})_{a b} (\sigma_{[3] \nu})_{c \dot d} \\
    (\sigma^\nu \tilde \sigma^{[5]})_{a \dot b}(\sigma_{[5]})_{c  d} &= -(\sigma^{[5]} \tilde \sigma^\nu )_{a \dot b}(\sigma_{[5]})_{c  d} + 10(\sigma^{[4]})_{a\dot  b} (\sigma_{[4] \nu})_{c  d}
\end{align}
Finally we present an explicit representation of the sigma matrices. The left handed matrices are given by
\begin{align}
(\sigma^{0})_{a b} &= \I_{2}\otimes \I_{2}\otimes \I_{2}\otimes \I_{2}
\\
(\sigma^{1})_{a b} &= \sigma^{2}\otimes \sigma^{2}\otimes \sigma^{2}\otimes \sigma^{2}
\\
(\sigma^{2})_{a b} &= \sigma^{2}\otimes \sigma^{2}\otimes \I_{2}\otimes \sigma^{1}
\\
(\sigma^{3})_{a b} &= \sigma^{2}\otimes \sigma^{2}\otimes \I_{2}\otimes \sigma^{3}
\\
(\sigma^{4})_{a b} &= \sigma^{2}\otimes \sigma^{1}\otimes \sigma^{2}\otimes \I_{2}
\\
(\sigma^{5})_{a b} &= \sigma^{2}\otimes \sigma^{3}\otimes \sigma^{2}\otimes \I_{2}
\\
(\sigma^{6})_{a b} &= \sigma^{2}\otimes \I_{2}\otimes \sigma^{1}\otimes \sigma^{2}
\\
(\sigma^{7})_{a b} &= \sigma^{2}\otimes \I_{2}\otimes \sigma^{3}\otimes \sigma^{2}
\\
(\sigma^{8})_{a b} &= \sigma^{1}\otimes \I_{2}\otimes \I_{2}\otimes \I_{2}
\\
(\sigma^{9})_{a b} &= \sigma^{3}\otimes \I_{2}\otimes \I_{2}\otimes \I_{2}.
\end{align}
and the right handed matrices by
\begin{align}
(\sigma^{0})^{a b} &= -\,\I_{2}\otimes \I_{2}\otimes \I_{2}\otimes \I_{2}\\
(\sigma^{1})^{a b} &= \sigma^{2}\otimes \sigma^{2}\otimes \sigma^{2}\otimes \sigma^{2}\\
(\sigma^{2})^{a b} &= \sigma^{2}\otimes \sigma^{2}\otimes \I_{2}\otimes \sigma^{1}\\
(\sigma^{3})^{a b} &= \sigma^{2}\otimes \sigma^{2}\otimes \I_{2}\otimes \sigma^{3}\\
(\sigma^{4})^{a b} &= \sigma^{2}\otimes \sigma^{1}\otimes \sigma^{2}\otimes \I_{2}\\
(\sigma^{5})^{a b} &= \sigma^{2}\otimes \sigma^{3}\otimes \sigma^{2}\otimes \I_{2}\\
(\sigma^{6})^{a b} &= \sigma^{2}\otimes \I_{2}\otimes \sigma^{1}\otimes \sigma^{2}\\
(\sigma^{7})^{a b} &= \sigma^{2}\otimes \I_{2}\otimes \sigma^{3}\otimes \sigma^{2}\\
(\sigma^{8})^{a b} &= \sigma^{1}\otimes \I_{2}\otimes \I_{2}\otimes \I_{2}. \\
(\sigma^{9})^{a b} &= \sigma^{3}\otimes \I_{2}\otimes \I_{2}\otimes \I_{2}.
\end{align}

\section{Detailed Superalgebra Closure Calculations}
The calculations for the bosons are relatively straightforward, and largely consist of recognizing gauge terms and equations of motion. For ease of interpretation throughout the following calculations, corresponding index (anti)symmetrizations are share a color.
\begin{align}
\begin{split}
    \{Q_a, Q_c\} h_{\mu \nu} &= D_{\textcolor{red}{(a|}}\left[(\sigma_{\textcolor{blue}{(\mu|}})_{\textcolor{red}{|c)}}{}^{\dot b} \psi_{\textcolor{blue}{|\nu)} \dot b}\right]\\
    &= (\sigma_{\textcolor{blue}{(\mu|}})_{\textcolor{red}{(c|}}{}^{\dot b} \Bigl[{id_0(\tilde \sigma^{\delta \rho})_{\dot{b} {\textcolor{red}{|a)}}}\partial_\delta h_{\rho
     \textcolor{blue}{|\nu)}}}_{\text{(A1)}} +{ ie_1 ( \tilde \sigma^{\delta \rho })_{\dot b {\textcolor{red}{|a)}}} \partial_\delta B_{\rho
     \textcolor{blue}{|\nu)}}}_{\text{(B1)}}+{ ie_2(\tilde \sigma_{\textcolor{blue}{|\nu)}}\sigma^{[3]})_{\dot b {\textcolor{red}{|a)}}}A_{[3]}}_{\text{(C1)}} \Bigr]
\end{split}
\end{align}
It is significantly easier to compute these terms individually. A good portion of the tensor algebra is omitted, but if a nontrivial manipulation is needed we will always mention it.

\begin{align}
\begin{split}
    (A1)= id_0(\sigma_{\textcolor{blue}{(\mu|}})_{\textcolor{red}{(c|}}{}^{\dot b} (\tilde \sigma^{\delta \rho})_{\dot{b} {\textcolor{red}{|a)}}}\partial_\delta h_{\rho
     \textcolor{blue}{|\nu)}}&= id_0(\sigma_{\textcolor{blue}{(\mu|}}{}^{\delta\rho})_{\textcolor{red}{(ac)}}\partial_{\delta}h_{\rho \textcolor{blue}{|\nu)}} + id_0(\sigma^{\rho})_{\textcolor{red}{(ac)}}\partial_{\textcolor{blue}{(\mu|}}h_{\rho \textcolor{blue}{|\nu)}} - id_0(\sigma^{\delta})_{\textcolor{red}{(ac)}}\partial_{\delta}h_{\textcolor{blue}{(\mu\nu)}}\\
     &= -4id_0\cancel{\partial}_{ac}h_{\mu\nu}+ \partial_{\textcolor{blue}{(\mu|}}\Bigl[2id_0(\sigma^{\rho})_{ac}h_{\rho \textcolor{blue}{|\nu)}}\Bigr]\\
     (B1) = ie_1(\sigma_{\textcolor{blue}{(\mu|}})_{\textcolor{red}{(c|}}{}^{\dot b} (\tilde \sigma^{\delta \rho})_{\dot{b} {\textcolor{red}{|a)}}}\partial_\delta B_{\rho
     \textcolor{blue}{|\nu)}}&= ie_1(\sigma_{\textcolor{blue}{(\mu|}}{}^{\delta\rho})_{\textcolor{red}{(ac)}}\partial_{\delta}B_{\rho \textcolor{blue}{|\nu)}} + ie_1(\sigma^{\rho})_{\textcolor{red}{(ac)}}\partial_{\textcolor{blue}{(\mu|}}B_{\rho \textcolor{blue}{|\nu)}} - ie_1(\sigma^{\delta})_{\textcolor{red}{(ac)}}\partial_{\delta}B_{\textcolor{blue}{(\mu\nu)}}\\
     &= \partial_{\textcolor{blue}{(\mu|}}\Bigl[2ie_1(\sigma^{\rho})_{ac}B_{\rho \textcolor{blue}{|\nu)}}\Bigr]\\
     (C1)=ie_2(\sigma_{\textcolor{blue}{(\mu|}})_{\textcolor{red}{(c|}}{}^{\dot b} (\tilde \sigma_{\textcolor{blue}{|\nu)}}\sigma^{[3]})_{\dot b {\textcolor{red}{|a)}}}A_{[3]} &= ie_2 {(\sigma_{\textcolor{blue}{(\mu\nu)}}\sigma^{[3]})_{\textcolor{red}{(ac)}} A_{[3]}}+ie_2(\eta_{\textcolor{blue}{(\mu\nu)}}\sigma^{[3]})_{\textcolor{red}{(ac)}}A_{[3]}\\
     & = 0
\end{split}
\end{align}
We are forced to set $d_0 = -1/2$ to satisfy superalgebra closure. Thus
\begin{equation} \{Q_a, Q_c\} h_{\mu \nu} = 2i\cancel{\partial}_{ac}h_{\mu\nu}+ \partial_{\textcolor{blue}{(\mu|}}\Bigl[-i(\sigma^{\rho})_{ac}h_{\rho|\textcolor{blue}{\nu)}}+ 2ie_1(\sigma^{\rho})_{ac}B_{\rho |\textcolor{blue}{\nu)}}\Bigr]\end{equation}
Calculation of closure on the 2-form proceeds similarly.
\begin{align}
\begin{split}
    \{Q_a, Q_c\} B_{\mu \nu} &= D_{\textcolor{red}{(a|}}\left[a_2(\sigma_{\textcolor{blue}{[\mu|}})_{\textcolor{red}{|c)}}{}^{\dot b} \psi_{\textcolor{blue}{|\nu]} \dot b} + (\sigma_{\mu \nu})_{\textcolor{red}{|c)}}{}^{b}\chi_b\right]\\
    &=a_2(\sigma_{\textcolor{blue}{[\mu|}})_{\textcolor{red}{(c|}}{}^{\dot b} \biggl[{\frac{i}{2}(\tilde \sigma^{\delta \rho})_{\dot{b} {\textcolor{red}{|a)}}}\partial_\delta h_{\rho
     \textcolor{blue}{|\nu]}} +ie_1 ( \tilde \sigma^{\delta \rho })_{\dot b {\textcolor{red}{|a)}}} \partial_\delta B_{\rho
     \textcolor{blue}{|\nu]}}}_{(A2)}+ {ie_2(\tilde \sigma_{\textcolor{blue}{|\nu]}}\sigma^{[3]})_{\dot b {\textcolor{red}{|a)}}}A_{[3]}}_{(B2)} \biggr]\\
     &\quad + (\sigma_{\mu \nu})_{\textcolor{red}{(a|}}{}^{b}\Bigl[{if_0 (\sigma^{\rho})_{b\textcolor{red}{|c)}} \partial_{\rho} \phi}_{(C2)} + {ie_0 (\sigma^{[3]})_{b \textcolor{red}{|c)}} A_{[3]}}_{(D2)}\Bigr]
\end{split}
\end{align}
The calculations for (A2) are almost identical to those for (A1) and (B1), with the replacement of some coefficients and a commutator instead of an anticommutator. The same steps lead to the equality
\begin{equation}(A2) = -4ia_2e_1\cancel{\partial}_{ac}B_{\mu\nu}+ \partial_{\textcolor{blue}{[\mu|}}\Bigl[2ia_2(\sigma^{\rho})_{ac}h_{\rho \textcolor{blue}{|\nu]}}+2ia_2e_1(\sigma^{\rho})_{ac}B_{\rho \textcolor{blue}{|\nu]}}\Bigr]\end{equation}
We compute the remaining terms individually.
\begin{align}
\begin{split}
    (B2) =ia_2e_2(\sigma_{\textcolor{blue}{[\mu|}})_{\textcolor{red}{(c|}}{}^{\dot b}(\tilde \sigma_{\textcolor{blue}{|\nu]}}\sigma^{[3]})_{\dot b {\textcolor{red}{|a)}}}A_{[3]}&= ia_2e_2(\sigma_{\textcolor{blue}{[\mu \nu]}}\sigma^{[3]})_{\textcolor{red}{(ac)}}A_{[3]} + ia_2e_2(\eta_{\textcolor{blue}{[\mu \nu]}}\sigma^{[3]})_{\textcolor{red}{(ac)}}A_{[3]}\\
    &=2ia_2e_2(\sigma_{\mu \nu}\sigma^{[3]})_{\textcolor{red}{(ac)}}A_{[3]}\\
    (C2)\:\ =\:\ if_0(\sigma_{\mu \nu})_{\textcolor{red}{(a|}}{}^{b} (\sigma^{\rho})_{b\textcolor{red}{|c)}} \partial_{\rho} \phi\:\:\ &=if_0(\sigma_{\mu \nu}{}^\rho)_{\textcolor{red}{(ac)}}\partial_\rho \phi+if_0(\sigma_{\textcolor{blue}{[\mu}}\delta^{\rho}_{ \textcolor{blue}{\nu]}})_{\textcolor{red}{(ac)}}\partial_\rho \phi \\
    &= \partial_{\textcolor{blue}{[\mu|}}\left[-2if_0(\sigma_{ \textcolor{blue}{|\nu]}})_{ac}\phi\right]\\
    (D2) \:\: =\:\ ie_0(\sigma_{\mu \nu})_{\textcolor{red}{(a|}}{}^{b}(\sigma^{[3]})_{b \textcolor{red}{|c)}} A_{[3]} \:\:  &= ie_0(\sigma_{\mu \nu} \sigma^{[3]})_{\textcolor{red}{(ac)}}A_{[3]}
\end{split}
\end{align}
We are forced to set $a_2e_1 = -1/2$ for algebra closure and $ e_0=-2a_2e_2$ for proper cancellation of extra terms. Thus
\begin{equation}\{Q_a, Q_c\}B_{\mu \nu} = 2i\cancel \partial_{ac}B_{\mu \nu}+ \partial_{\textcolor{blue}{[\mu|}}\left[-i(\sigma^{\rho})_{ac}B_{\rho |\textcolor{blue}{\nu]}} + 2ia_2(\sigma^\rho)_{ac}h_{\rho |\textcolor{blue}{\nu]}} - 2if_0(\sigma_{|\textcolor{blue}{\nu]}})_{ac}\phi\right]\end{equation}
Finally, the calculation for the dilatino.
\begin{align}
\begin{split}
    \{Q_a, Q_c\} \phi &= D_{\textcolor{red}{(a|}}[c_1 \chi_{\textcolor{red}{|c)}}]\\
    &= ic_1 f_0 (\sigma^{[1]} )_{\textcolor{red}{(ac)}}\partial_{[1]}\phi + ic_1e_0 (\sigma^{[3]})_{\textcolor{red}{(ac)}}A_{[3]}\\
    &= 2ic_1f_0 \cancel \partial_{ac}\phi
\end{split}
\end{align}
Closure requires us to set $c_1f_0 = 1$ and thus
\begin{equation}\{Q_a, Q_c\}\phi = 2i\cancel \partial_{ac}\phi\end{equation}

While the omitted tensor manipulations in this section are still straightforward, they require more algebra than those for the bosons. More detailed calculations are available upon request.

Recall that $A_{\mu \nu \rho} = \frac{1}{3!}\partial_{[\mu }B_{\nu \rho]}$. If $X^{U \mu \nu \rho}_L$ is some tensor with arbitrary sets of upper and lower indices $U$ and $L$ and $X^{U \mu \nu \rho}_L$ is totally antisymmetric in $\mu \nu \rho$, then $X^{U \mu \nu \rho}_L A_{\mu \nu \rho} = X^{U \mu \nu \rho}_L \partial_\mu B_{\nu \rho}$. We will use this removal of the antisymmetrization of $A_{\mu \nu \rho}$ (implicitly) many times in the following calculations.

From closure on the bosons we set $d_0 = -1/2$. Therefore
\begin{align}
\begin{split}
    \{Q_a, Q_c\}\psi_{\mu \dot b} &= D_{\textcolor{red}{(a|}}\left[-\frac{i}{2}(\tilde \sigma)^{\nu \rho})_{\dot b \textcolor{red}{|c)}}\partial_\nu h_{\rho \mu}+ ie_1(\tilde \sigma^{\nu \rho})_{\dot b \textcolor{red}{|c)}} \partial_\nu B_{\rho \mu} + i e_2 (\tilde \sigma_\mu \sigma^{[3]})_{\dot b \textcolor{red}{|c)}} A_{[3]} \right]\\
    &= -\frac{i}{2}(\tilde \sigma)^{\nu \rho})_{\dot b \textcolor{red}{(c|}}(\sigma_{\textcolor{blue}{(\rho|}})_{\textcolor{red}{|a)}}{}^{\dot d}\partial_\nu\psi_{\textcolor{blue}{|\mu)}\dot d}+ie_1(\tilde \sigma^{\nu \rho})_{\dot b \textcolor{red}{(c|}}\left[a_2(\sigma_{\textcolor{blue}{[\rho|}})_{\textcolor{red}{|a)}}{}^{\dot d}\partial_\nu\psi_{\textcolor{blue}{|\mu]}\dot d}+(\sigma_{\rho \mu})_{\textcolor{red}{|a)}}{}^{ d}\partial_\nu\chi_d\right]\\
    &\quad +ie_2(\tilde \sigma_\mu \sigma^{\nu \rho \delta})_{\dot b \textcolor{red}{(c|}}\left[a_2(\sigma_{\textcolor{blue}{[\rho|}})_{\textcolor{red}{|a)}}{}^{\dot d}\partial_\nu\psi_{\textcolor{blue}{|\delta]}\dot d}+(\sigma_{\rho \delta})_{\textcolor{red}{|a)}}{}^{ d}\partial_\nu\chi_d\right]\\
    &=-{i(\tilde \sigma^{\nu \rho})_{\dot b \textcolor{red}{(c|}}(\sigma_{\rho})_{\textcolor{red}{|a)}}{}^{\dot d}\partial_\nu\psi_{\mu\dot d}}_{(A3)}+{ie_1(\tilde \sigma^{\nu \rho})_{\dot b \textcolor{red}{(c|}}(\sigma_{\rho \mu})_{\textcolor{red}{|a)}}{}^{ d}\partial_\nu\chi_d}_{(B3)}\\
    &\quad+{ie_2(\tilde \sigma_\mu \sigma^{\nu \rho \delta})_{\dot b \textcolor{red}{(c|}}(\sigma_{\rho \delta})_{\textcolor{red}{|a)}}{}^{ d}\partial_\nu\chi_d}_{(C3)}+ {ie_2a_2(\tilde \sigma_\mu \sigma^{\nu \rho \delta})_{\dot b \textcolor{red}{(c|}}(\sigma_{\textcolor{blue}{[\rho|}})_{\textcolor{red}{|a)}}{}^{\dot d}\partial_\nu\psi_{\textcolor{blue}{|\delta]}\dot d}}_{(D3)}
\end{split}
\end{align}
The Fierz identity (A.12) found in Appendix A is crucial to the following calculations. We'll illustrate the procedure for each of these terms by appropriately expanding (A3), and give the results for the other terms.
\begin{align}
\begin{split}
    (A3)  &= - i(\tilde \sigma^{\nu \rho})_{\dot b \textcolor{red}{(c|}}(\sigma_{\rho})_{\textcolor{red}{|a)}}{}^{\dot d}\partial_\nu\psi_{\mu\dot d} \\&= -i(\tilde \sigma^{\nu \rho})_{\dot b e}\left[\delta^e_{\textcolor{red}{(c|}}\delta^f_{\textcolor{red}{|a)}}\right](\sigma_{\rho})_{f}{}^{\dot d}\partial_\nu\psi_{\mu\dot d} \\
    &= -i(\tilde \sigma^{\nu \rho})_{\dot b e}\left[\frac{1}{8}(\sigma_{[1]})_{ac}(\tilde \sigma^{[1]})^{ef} + \frac{1}{16 \times 5!}(\sigma_{[5]})_{ac}(\tilde \sigma^{[5]})^{ef}\right](\sigma_{\rho})_{f}{}^{\dot d}\partial_\nu\psi_{\mu\dot d}\\
    &=\left[-\frac{i}{8}(\sigma_{[1]})_{ac}(\tilde \sigma^{\nu \rho} \tilde \sigma^{[1]} \sigma_\rho)_{\dot b}{}^{\dot d}-\frac{i}{16 \times 5!}(\sigma_{[5]})_{ac}(\tilde \sigma^{\nu \rho}\tilde \sigma^{[5]}\sigma_\rho)_{\dot b}{}^{\dot d}\right]\partial_\nu\psi_{\mu\dot d}\\
    &=2i\cancel \partial_{ac}\psi_{\mu \dot b} - \frac{7i}{8}(\sigma_{[1]})_{ac}(\tilde \sigma^{[1]} \sigma^\nu)_{\dot b}{}^{\dot d}\partial_\nu \psi_{\mu \dot d}+\frac{i}{120}(\sigma_{[5]})_{ac}(\tilde \sigma^{[5]} \sigma^\nu)_{\dot b}{}^{\dot d}\partial_\nu \psi_{\mu \dot d}\\
    &\quad + \partial_\mu \left[\frac{7i}{8}(\sigma_{[1]})_{ac}(\tilde \sigma^{[1]} \sigma^\nu)_{\dot b}{}^{\dot d}\psi_{\nu \dot d}-\frac{i}{120}(\sigma_{[5]})_{ac}(\tilde \sigma^{[5]} \sigma^\nu)_{\dot b}{}^{\dot d} \psi_{\nu \dot d}\right]
\end{split}
\end{align}
Proceeding in the same manner, we find that
\begin{align}
\begin{split}
    (B3) &= ie_1(\tilde \sigma^{\nu \rho})_{\dot b \textcolor{red}{(c|}}(\sigma_{\rho \mu})_{\textcolor{red}{|a)}}{}^{ d}\partial_\nu\chi_d \\
    &= ie_1\left[\frac{1}{8}(\sigma_{[1]})_{ac}(\tilde \sigma^{\nu \rho}\tilde \sigma^{[1]} \sigma_{\rho \mu})_{\dot b}{}^{ d} + \frac{1}{16 \times 5!}(\sigma_{[5]})_{ac}(\tilde \sigma^{\nu \rho}\tilde \sigma^{[5]} \sigma_{\rho \mu})_{\dot b}{}^{ d}\right]\partial_\nu \chi_d\\
    &= ie_1\biggl[\frac{3}{4}(\sigma_{[1]})_{ac}(\tilde\sigma_\mu \sigma^{[1]} \tilde \sigma^\nu)_{\dot b}{}^d -\fbox{$\frac{7}{4}(\sigma^\nu)_{ac}(\tilde \sigma_\mu)_{\dot b}{}^d$} \\
    &\qquad\qquad- \frac{7}{4}(\sigma_\mu)_{ac}(\tilde \sigma^\nu)_{\dot b}{}^d - \frac{1}{8 \times 5!}(\sigma_{[5]})_{ac}(\tilde \sigma_\mu\sigma^{[5]}\tilde \sigma^\nu)_{\dot b}{}^d\\
    &\qquad\qquad+\frac{1}{8 \times 4!}(\sigma_{[4]\mu})_{ac} (\tilde \sigma^{[4]}\tilde \sigma^\nu)_{\dot b}{}^d +\fbox{$ \frac{1}{8 \times 4!}(\sigma_{[4]}{}^\nu)_{ac}(\tilde \sigma_\mu \sigma^{[4]})_{\dot b}{}^d$} \biggr]\partial_\nu \chi_d\\
    &\quad + ie_1\partial_\mu \left[\frac{13}{8}(\sigma_{[1]})_{ac}(\tilde \sigma^{[1]})_{\dot b}{}^d \chi_d - \frac{11}{16 \times 5!}(\sigma_{[5]})_{ac}(\tilde \sigma^{[5]})_{\dot b}{}^d \chi_d\right]
\end{split}
\end{align}
\begin{align}
\begin{split}
    (C3) &= ie_2(\tilde \sigma_\mu \sigma^{\nu \rho \delta})_{\dot b \textcolor{red}{(c|}}(\sigma_{\rho \delta})_{\textcolor{red}{|a)}}{}^{ d}\partial_\nu\chi_d \\&= ie_2 \left[\frac{1}{8}(\sigma_{[1]})_{ac}(\tilde \sigma_\mu \sigma^{\nu \rho \delta}\tilde \sigma^{[1]}\sigma_{\rho \delta})_{\dot b}{}^d + \frac{1}{16 \times 5!}(\sigma_{[5]})_{ac}(\tilde \sigma_\mu \sigma^{\nu \rho \delta}\tilde \sigma^{[5]}\sigma_{\rho \delta})_{\dot b}{}^d\right] \partial_\nu \chi_d\\
    &= ie_2\biggl[5(\sigma_{[1]})_{ac}(\tilde \sigma_\mu \sigma^{[1]}\tilde \sigma^\nu)_{\dot b}{}^d - \fbox{$14(\sigma^\nu)_{ac}(\tilde \sigma_\mu)_{\dot b}{}^d$} \\
    &\qquad\qquad - \frac{1}{2 \times 5!}(\sigma_{[5]})_{ac}(\tilde \sigma_\mu \sigma^{[5]} \tilde \sigma^\nu)_{\dot b}{}^d +\fbox{$ \frac{1}{4!}(\sigma_{[4]}{}^\nu )_{ac}(\tilde \sigma_\mu \sigma^{[4]})_{\dot b}{}^d$}\biggr] \partial_\nu \chi_d
\end{split}
\end{align}
\begin{align}
\begin{split}
    (D3) &= ie_2a_2(\tilde \sigma_\mu \sigma^{\nu \rho \delta})_{\dot b \textcolor{red}{(c|}}(\sigma_{\textcolor{blue}{[\rho|}})_{\textcolor{red}{|a)}}{}^{\dot d}\partial_\nu\psi_{\textcolor{blue}{|\delta]}\dot d}\\
    &= ie_2a_2\left[\frac{1}{4}(\sigma_{[1]})_{ac}(\tilde \sigma_\mu \sigma^{\nu \rho \delta}\tilde \sigma^{[1]}\sigma_\rho)_{\dot b}{}^{\dot d} + \frac{1}{8 \times 5!}(\sigma_{[5]})_{ac}(\tilde \sigma_\mu \sigma^{\nu \rho \delta}\tilde \sigma^{[5]}\sigma_\rho)_{\dot b}{}^{\dot d} \right]\partial_\nu \psi_{\delta \dot d}\\
    &= ie_2a_2\biggl[-(\sigma_{[1]})_{ac}(\tilde \sigma_{\mu}\sigma^{\textcolor{blue}{[\nu|}}\tilde \sigma^{[1]}\sigma^{\textcolor{blue}{|\delta]}})_{\dot b}{}^{\dot d}- 2(\sigma^{\textcolor{blue}{[\nu|}})_{ac}(\tilde \sigma_\mu \sigma^{\textcolor{blue}{|\delta]}})_{\dot b}{}^{\dot d}\\
    &\qquad\qquad +\frac{1}{2}(\sigma_{[1]})_{ac}(\tilde \sigma_\mu \sigma^{\nu [1]\delta})_{\dot b}{}^{\dot d}+\frac{1}{8 \times 5!}(\sigma_{[5]})_{ac}(\tilde \sigma_\mu  \sigma^{\textcolor{blue}{[\nu|}} \tilde \sigma^{[5]} \sigma^{\textcolor{blue}{|\delta]}})_{\dot b}{}^{\dot d}
    \biggr]\partial_\nu \psi_{\delta \dot d}
\end{split}
\end{align}
The boxed terms are the remaining terms that do not contribute appropriately to closure. They vanish as long as we set $e_2 = -e_1/8$, which gives us the final equation for closure on the gravitino:
\begin{align}
    \{Q_a, Q_c\}\psi_{\mu \dot b} &= 2i\cancel \partial_{ac}\psi_{\mu \dot b} - \frac{7i}{8}(\sigma_{[1]})_{ac}(\tilde \sigma^{[1]} \sigma^\nu)_{\dot b}{}^{\dot d}\partial_\nu \psi_{\mu \dot d}+\frac{i}{120}(\sigma_{[5]})_{ac}(\tilde \sigma^{[5]} \sigma^\nu)_{\dot b}{}^{\dot d}\partial_\nu \psi_{\mu \dot d}\\
    &\quad\tag*{} +ie_1\biggl[\frac{1}{8}(\sigma_{[1]})_{ac}(\tilde\sigma_\mu \sigma^{[1]} \tilde \sigma^\nu)_{\dot b}{}^d - \frac{7}{4}(\sigma_\mu)_{ac}(\tilde \sigma^\nu)_{\dot b}{}^d \\
    &\qquad \qquad\tag*{}- \frac{1}{16 \times 5!}(\sigma_{[5]})_{ac}(\tilde \sigma_\mu\sigma^{[5]}\tilde \sigma^\nu)_{\dot b}{}^d+\frac{1}{8 \times 4!}(\sigma_{[4]\mu})_{ac} (\tilde \sigma^{[4]}\tilde \sigma^\nu)_{\dot b}{}^d \biggr]\partial_\nu \chi_d\\
    &\quad\tag*{} - \frac{ie_1a_2}{8}\biggl[-(\sigma_{[1]})_{ac}(\tilde \sigma_{\mu}\sigma^{\textcolor{blue}{[\nu|}}\tilde \sigma^{[1]}\sigma^{\textcolor{blue}{|\delta]}})_{\dot b}{}^{\dot d}- 2(\sigma^{\textcolor{blue}{[\nu|}})_{ac}(\tilde \sigma_\mu \sigma^{\textcolor{blue}{|\delta]}})_{\dot b}{}^{\dot d}\\
    &\qquad\qquad\tag*{} +\frac{1}{2}(\sigma_{[1]})_{ac}(\tilde \sigma_\mu \sigma^{\nu [1]\delta})_{\dot b}{}^{\dot d}+\frac{1}{8 \times 5!}(\sigma_{[5]})_{ac}(\tilde \sigma_\mu  \sigma^{\textcolor{blue}{[\nu|}} \tilde \sigma^{[5]} \sigma^{\textcolor{blue}{|\delta]}})_{\dot b}{}^{\dot d}
    \biggr]\partial_\nu \psi_{\delta \dot d}\\
    &\quad\tag*{} + ie_1 \left[\frac{13}{8}(\sigma_{[1]})_{ac}(\tilde \sigma^{[1]})_{\dot b}{}^d  - \frac{11}{16 \times 5!}(\sigma_{[5]})_{ac}(\tilde \sigma^{[5]})_{\dot b}{}^d\right] \partial_\mu \chi_d\\
     &\quad\tag*{} +\partial_\mu \left[\frac{7i}{8}(\sigma_{[1]})_{ac}(\tilde \sigma^{[1]} \sigma^\nu)_{\dot b}{}^{\dot d}-\frac{i}{120}(\sigma_{[5]})_{ac}(\tilde \sigma^{[5]} \sigma^\nu)_{\dot b}{}^{\dot d} \right]   \psi_{\nu \dot d}
\end{align}
Let's move onto the calculation for the dilatino. From closure on the bosons, $f_0 c_1 = 1$.
\begin{align}
\begin{split}
    \{Q_a, Q_c\} \chi_b &= D_{\textcolor{red}{(a|}}\left[if_0 (\sigma^\nu)_{b \textcolor{red}{|c)}}\partial_\nu \phi + ie_0(\sigma^{[3]})_{b \textcolor{red}{|c)}}A_{[3]}\right]\\
    &= {i(\sigma^\nu)_{b \textcolor{red}{(c|}}\partial_\nu\chi_{\textcolor{red}{|a)}}}_{(A4)} + {ie_0(\sigma^{\nu\rho \delta})_{b \textcolor{red}{(c|}}(\sigma_{\rho \delta})_{\textcolor{red}{|a)}}{}^d \partial_\nu\chi_d}_{(B4)} + {ie_0a_2(\sigma^{\nu\rho \delta})_{b \textcolor{red}{(c|}}(\sigma_{\textcolor{blue}{[\rho|}})_{\textcolor{red}{|a)}}{}^{\dot d}\partial_{\textcolor{blue}{|\delta]}}\psi_{\delta \dot d}}_{(C4)}
\end{split}
\end{align}
Each term is calculated in the same manner as above, by applying (A.16) and simplifying.
\begin{align}
\begin{split}
    (A4) &= i(\sigma^\nu)_{b \textcolor{red}{(c|}}\partial_\nu\chi_{\textcolor{red}{|a)}}\\
    &=i\left[\frac{1}{8}(\sigma_{[1]})_{ac}(\sigma^\nu\tilde \sigma^{[1]})_b{}^d + \frac{1}{16 \times 5!}(\sigma_{[5]})_{ac}(\sigma^\nu \tilde \sigma^{[5]})_b{}^d\right]\partial_\nu \chi_d\\
    &= \frac{i}{4}\cancel \partial_{ac}\chi_b + i \biggl[\: \fbox{$\frac{1}{8 \times 4!}(\sigma_{[4]}{}^\nu)_{ac}(\sigma^{[4]})_b{}^d$}\\
    &\quad - \frac{1}{8}(\sigma_{[1]})_{ac}(\sigma^{[1]}\tilde \sigma^\nu)_b{}^d - \frac{1}{16 \times 5!}(\sigma_{[5]})_{ac}(\sigma^{[5]}\tilde \sigma^\nu)_b{}^d\biggr]\partial_\nu \chi_d
\end{split}
\end{align}
\begin{align}
\begin{split}
    (B4) &=ie_0(\sigma^{\nu\rho \delta})_{b \textcolor{red}{(c|}}(\sigma_{\rho \delta})_{\textcolor{red}{|a)}}{}^d \partial_\nu\chi_d\\
    &= ie_0 \left[\frac{1}{8}(\sigma_{[1]})_{ac}(\sigma^{\nu \rho \delta} \tilde \sigma^{[1]}\sigma_{\rho \delta})_b{}^d + \frac{1}{16 \times 5!}(\sigma_{[5]})_{ac}(\sigma^{\nu \rho \delta} \tilde \sigma^{[5]}\sigma_{\rho \delta})_b{}^d \right]\partial_\nu \chi_d\\
    &= -14ie_0\cancel \partial_{ac}\chi_b + ie_0\biggl[\: \fbox{$\frac{1}{4!}(\sigma_{[4]}{}^\nu)_{ac}(\sigma^{[4]})_b{}^d$}\\
    &\quad + 5(\sigma_{[1]})_{ac}(\sigma^{[1]}\tilde \sigma^\nu)_b{}^d - \frac{1}{2 \times 5!}(\sigma_{[5]})_{ac}(\sigma^{[5]}\tilde \sigma^\nu)_b{}^d\biggr]\partial_\nu \chi_d
\end{split}
\end{align}
\begin{align}
\begin{split}
    (C4) &= ie_0a_2(\sigma^{\nu\rho \delta})_{b \textcolor{red}{(c|}}(\sigma_{\textcolor{blue}{[\rho|}})_{\textcolor{red}{|a)}}{}^{\dot d}\partial_{\textcolor{blue}{|\delta]}}\psi_{\delta \dot d}\\
    &=ie_0a_2\left[\frac{1}{4}(\sigma_{[1]})_{ac}(\sigma^{\nu \rho \delta}\tilde \sigma^{[1]}\sigma_\rho)_b{}^{\dot d}+\frac{1}{8 \times 5!}(\sigma_{[5]})_{ac}(\sigma^{\nu \rho \delta}\tilde \sigma^{[5]}\sigma_\rho)_b{}^{\dot d}\right]\partial_\nu \psi_{\delta \dot d}\\
    &= ie_0a_2 \biggl[\frac{3}{2}(\sigma_{[1]})_{ac}(\sigma^{\nu \delta [1]})_b{}^{\dot d}+2(\sigma^{\textcolor{blue}{[\delta|}})_{ac}(\sigma^{\textcolor{blue}{|\nu]}})_b{}^{\dot d}+ \frac{1}{8 \times 5!}(\sigma_{[5]})_{ac}(\sigma^{\textcolor{blue}{[\nu|}}\tilde \sigma^{[5]}\sigma^{\textcolor{blue}{|\delta]}})_{b}{}^{\dot d}\biggr]\partial_\nu \psi_{\delta \dot d}
\end{split}
\end{align}
The boxed terms are again the only ones that do not contribute appropriately, so we are forced to set $e_0 = -1/8 $. Altogether we conclude
\begin{align}
    \{Q_a, Q_c\}\chi_b &= 2i\cancel \partial_{ac}\chi_b -\frac{3i}{4}(\sigma_{[1]})_{ac}(\sigma^{[1]}\tilde \sigma^\nu)_{b}{}^d\partial_\nu \chi_d  +  ie_0a_2 \biggl[\frac{3}{2}(\sigma_{[1]})_{ac}(\sigma^{\nu \delta [1]})_b{}^{\dot d}\\
    &\quad\tag*{}+2(\sigma^{\textcolor{blue}{[\delta|}})_{ac}(\sigma^{\textcolor{blue}{|\nu]}})_b{}^{\dot d}+ \frac{1}{8 \times 5!}(\sigma_{[5]})_{ac}(\sigma^{\textcolor{blue}{[\nu|}}\tilde \sigma^{[5]}\sigma^{\textcolor{blue}{|\delta]}})_{b}{}^{\dot d}\biggr]\partial_\nu \psi_{\delta \dot d}
\end{align}
To recap, closure of the superalgebra produces the five constraint equations
\begin{equation}d_0 = -1/2, \qquad a_2e_1 = -1/2, \qquad c_1f_0 = 1, \qquad e_2 = -e_1/8, \qquad e_0 = -1/8.\end{equation}

\section{Detailed Supercurrent Calculations}
Imposing Lorentz invariance, reality, and engineering-dimension constraints, the most general quadratic free-field Lagrangian built from our fields takes the form
\begin{equation}\mathcal L = a_3 \mathcal{R}_0 + ib_3 \psi_{\mu}{}^a (\sigma^{\mu\nu\rho})_a{}^{\dot b} \partial_{\nu} \psi_{\rho \dot b} + c_3 A_{[3]}A^{[3]}+ id_3  \chi^{\dot c}(\tilde{\sigma}^\mu)_{ \dot c}{}^{b} \partial_{\mu} \chi_b + e_3\partial_{\mu} \phi \partial^{\mu} \phi\end{equation}
where $a_3, b_3, c_3, d_3, e_3$ are a priori unknown real coefficients. At the free-field level, overall normalizations can be shifted by field rescalings; thus the key physical requirement is that each kinetic term have the canonical sign and normalization consistent with our metric signature and reality conditions. Choosing $a_3 = -2$, $c_3 = -3/2$ and $e_3 = -1/2$ places the bosonic sector in standard canonical form.
Calculation of the supercurrent will produce the remaining constraints needed to uniquely identify the coefficients of our superalgebra and Lagrangian. Recall that the supercurrent is found from the conserved current equation
\begin{align}
\partial_\nu (J^\nu)_a = \partial_\nu \left[\sum_i \frac{\partial \mathcal L}{\partial (\partial_\nu \Phi_i)} Q_a \Phi_i \right]-Q_a \mathcal L
\end{align}
We will first focus on the boundary piece $Q_a \mathcal L$ and then the Noether piece, in the same manner as for the closure calculations, starting with the bosonic terms.
We will compute the action of our supercharge on each of the terms in the Lagrangian individually, and then combine them to determine our constraints. The omitted algebra here is slightly more difficult than above, but only to the extent that integration by parts is applied several times implicitly. We begin by calculating the action on the Ricci scalar $\mathcal{R} = - 2\mathcal{R}_0$.
\begin{align}
\begin{split}
    Q_a \mathcal{R} &= -\partial_{\rho}Q_a h_{\mu\nu}\partial^{\rho}h^{\mu\nu} + \partial^\rho Q_a h\partial_\rho h -\partial^\rho Q_a h \partial^\mu h_{\rho\mu} \\
    &\quad-\partial^\rho  h \partial^\mu Q_a h_{\rho\mu} + 2  \partial^{\mu} Q_a h_{\mu \nu} \partial_\rho h^{\rho \nu} \\
    &= -2(\sigma_{\mu})_a{}^{\dot b}\partial_{\rho} \psi_{\nu \dot b}\partial^{\rho}h^{\mu\nu} + 2 (\sigma^{\nu})_a{}^{\dot b}\partial^\rho \psi_{\nu \dot b}\partial_\rho h -2  (\sigma^{\nu})_a{}^{\dot b}\partial^\rho \psi_{\nu \dot b} \partial^\mu h_{\rho\mu}\\
    &\quad -\partial^\rho  h \partial^\mu (\sigma_{\textcolor{blue}{(\rho|}})_a{}^{\dot b} \psi_{\textcolor{blue}{|\mu)} \dot b} + 2  (\sigma_{\textcolor{blue}{(\mu|}})_a{}^{\dot b}\partial^{\mu}  \psi_{\textcolor{blue}{\nu)} \dot b} \partial_\rho h^{\rho \nu}
\end{split}
\end{align}
Next we compute the field strength term.
\begin{align}
\begin{split}
      - \frac{3}{2} Q_a \bigl{[} A_{\mu \nu \rho}A^{\mu \nu \rho} \bigr{]} &= - 3 \bigl[Q_a A_{\mu \nu \rho}\bigr]A^{\mu \nu \rho} \\
    &= -\partial_\mu \big{[}  a_2(\sigma_{\textcolor{blue}{[\nu|}})_a{}^{\dot b} \psi_{\textcolor{blue}{|\rho]} \dot b} + (\sigma_{\nu \rho})_a{}^b \chi_b
    \big{]}\partial^\mu B^{\nu \rho} \\
    &\quad- 2\partial_\rho \big{[} a_2(\sigma_{\textcolor{blue}{[\mu|}})_{a}{}^{\dot b} \psi_{\textcolor{blue}{|\nu]} \dot b} + (\sigma_{\mu \nu})_a{}^b \chi_b
    \big{]} \partial^\mu B^{\nu \rho}\\
    &=  - 2a_2(\sigma_{\nu})_a{}^{\dot b} \partial_\mu  \psi_{\rho \dot b}  \partial^\mu B^{\nu \rho}  - (\sigma_{\nu \rho})_a{}^b \partial_\mu  \chi_b
    \partial^\mu B^{\nu \rho}\\
    &\quad- 2a_2\partial_\rho (\sigma_{\textcolor{blue}{[\mu|}})_{a}{}^{\dot b} \psi_{\textcolor{blue}{|\nu]} \dot b}  \partial^\mu B^{\nu \rho} - 2 (\sigma_{\mu \nu})_a{}^b \partial_\rho \chi_b
     \partial^\mu B^{\nu \rho}
\end{split}
\end{align}
Finally we act on the dilaton terms.
\begin{align}
\begin{split}
    - \frac{1}{2} Q_a\big{[} \partial_{\mu} \phi \partial^{\mu} \phi \big{]} &= -\bigl[\partial_{\mu} Q_a  \phi\bigr] \partial^{\mu} \phi = -c_1\partial_\mu \chi_a \partial^\mu \phi
\end{split}
\end{align}
As with the closure calculation, the fermions are significantly more difficult to work with. Recall from the superalgebra closure that $e_2 = -e_1/8$. Then
\begin{align}
\begin{split}
    ib_3 Q_a \big{[} \psi_{\mu}{}^c (\sigma^{\mu\nu\rho})_c{}^{\dot b} \partial_{\nu} \psi_{\rho \dot b} \big{]} &=  ib_3 Q_a \psi_{\mu}{}^c (\sigma^{\mu\nu\rho})_c{}^{\dot b} \partial_{\nu} \psi_{\rho \dot b} -   ib_3 \psi_{\mu}{}^c (\sigma^{\mu\nu\rho})_c{}^{\dot b} \partial_{\nu} Q_a \psi_{\rho \dot b} \\
    &= 2ib_3 Q_a \psi_{\mu}{}^c (\sigma^{\mu\nu\rho})_c{}^{\dot b} \partial_{\nu} \psi_{\rho \dot b} - ib_3 \partial_\nu \big{[} \psi_{\mu}{}^c (\sigma^{\mu\nu\rho})_c{}^{\dot b}  Q_a \psi_{\rho \dot b} \big{]} \\
    &= 2 ib_3 \left[ -\frac{i}{2}(\tilde \sigma^{\delta \xi})^c{}_{ a}\partial_\delta h_{\xi
     \mu} +ie_1( \tilde \sigma^{\delta \xi })^c{}_{ a} \partial_\delta B_{\xi\mu} -\frac{ie_1}{8}(\tilde \sigma_\mu\sigma^{[3]})^c{}_{ a}A_{[3]}
    \right](\sigma^{\mu \nu \rho})_c{}^{\dot b}   \partial_{\nu} \psi_{\rho \dot b}\\
    &\quad - ib_3 \partial_\nu \big{[} \psi_{\mu \dot b} (\sigma^{\mu\nu\rho})_c{}^{\dot b}  Q_a \psi_{\rho \dot b} \big{]} \\
    &= (\sigma^{\mu \nu \rho})^{\dot b c}\biggl{[}- \underbrace{b_3(\tilde \sigma^{\delta \xi})_{c a}\partial_\delta h_{\xi
     \mu}}_\text{(A5)}  +\underbrace{2b_3e_1 ( \tilde \sigma^{\delta \xi })_{c a} \partial_\delta B_{\xi\mu}}_\text{(B5)} \\
     &\qquad \qquad \quad-\underbrace{ \frac{b_3e_1}{4}(\tilde \sigma_\mu\sigma^{[3]})_{c a}A_{[3]}}_\text{(C5)}
    \biggr{]}  \partial_{\nu} \psi_{\rho \dot b} \\
    &\quad - ib_3 \partial_\nu \big{[} \psi_{\mu}{}^c (\sigma^{\mu\nu\rho})_c{}^{\dot b}  Q_a \psi_{\rho \dot b} \big{]}
\end{split}
\end{align}
We will show the explicit calculation for (A5) and show results for the remaining terms.
\begin{align}
\begin{split}
    (A5) &= -b_3 (\sigma^{\mu \nu \rho}\tilde \sigma^{\delta \xi})^{\dot b}{}_a\partial_\delta h_{\xi
     \mu}  \partial_{\nu} \psi_{\rho \dot b}\\
     &= b_3 \biggl{[} -\eta^{\mu\delta }(\sigma^{ \nu \rho \xi})^{\dot b}{}_{ a}  + \eta^{ \mu\xi}(\sigma^{\nu \rho \delta})^{\dot b}{}_{ a}  - \eta^{\nu\xi }(\sigma^{\mu \rho \delta})^{\dot b}{}_{ a}   \\
     &\quad+ \eta^{\rho \xi }(\sigma^{\mu\nu  \delta})^{\dot b}{}_{ a}  - \eta^{\textcolor{blue}{[\rho|} \delta} \eta^{\textcolor{blue}{|\nu|} \xi}(\sigma^{\textcolor{blue}{|\mu]}})^{\dot b}{}_{ a}
     \biggr{]}\partial_\delta h_{\xi \mu}    \partial_{\nu} \psi_{\rho \dot b} \\
    &= -b_3 \partial_\nu \left[(\sigma^{\mu\nu  \delta})_a{}^{\dot b}\partial_\delta h^\rho_\mu \psi_{\rho \dot b}  +(\sigma^{\nu \rho \delta})_a{}^{\dot b}\partial_\delta h \psi_{\rho \dot b}\right]\\
    &\quad +b_3 \left[(\sigma^{\nu \rho \xi})_a{}^{\dot b}\partial^\mu h_{\mu \xi} \partial_\nu \psi_{\rho \dot b} + \fbox{$(\sigma^{\mu \rho \delta})_a{}^{\dot b}\partial_\delta h_{\mu \nu}\partial^\nu \psi_{\rho \dot b}$}\right]\\
    &\quad -b_3\eta^{\textcolor{blue}{[\rho|} \delta} \eta^{\textcolor{blue}{|\nu|} \xi}(\sigma^{\textcolor{blue}{|\mu]}})^{\dot b}{}_{ a}
     \partial_\delta h_{\xi \mu}    \partial_{\nu} \psi_{\rho \dot b}\\
    &= b_3 \partial_\nu \biggl[-(\sigma^{\mu\nu  \delta})_a{}^{\dot b}\partial_\delta h^\rho_\mu \psi_{\rho \dot b}  -(\sigma^{\nu \rho \delta})_a{}^{\dot b}\partial_\delta h \psi_{\rho \dot b}\\
    &\qquad \qquad+  (\sigma^{\mu \rho \delta})_a{}^{\dot b}\partial_\delta h_{\mu }^\nu\psi_{\rho \dot b}+(\sigma^{\nu \rho \delta})_a{}^{\dot b}\partial^\mu h_{\mu \delta}\psi_{\rho \dot b}\biggr]\\
    &\quad -b_3\eta^{\textcolor{blue}{[\rho|} \delta} \eta^{\textcolor{blue}{|\nu|} \xi}(\sigma^{\textcolor{blue}{|\mu]}})_a{}^{\dot b}
     \partial_\delta h_{\xi \mu}    \partial_{\nu} \psi_{\rho \dot b}\\
\end{split}
\end{align}

\noindent To illustrate some of the suppressed algebra, the final equality above follows by applying integration by parts twice to the boxed term. Using symmetry of the graviton, this term cancels with its bracketed partner, leaving only the total derivative. Similarly for the remaining terms,
\begin{align}
\begin{split}
    (B5) &= 2b_3e_1(\sigma^{\mu \nu \rho}\tilde \sigma^{\delta \xi})^{\dot b}{}_a \partial_\delta B_{\xi \mu} \partial_\nu \psi_{\rho \dot b}\\
    &= 2b_3e_1 \partial_\nu\biggl[(\sigma^{\mu \nu \rho \delta \xi})_{a}{}^{\dot b}\partial_\delta B_{\xi \mu} \psi_{\rho \dot b} + (\sigma^{\mu \nu \delta})_a{}^{\dot b}\partial_\delta B^\rho{}_{ \mu}\psi_{\rho \dot b}\biggr]\\
    &\quad +2b_3e_1\biggl[-(\sigma^{\mu \rho \delta})_a{}^{\dot b}\partial_\delta B^\nu{}_\mu \partial_\nu\psi_{\rho \dot b} - (\sigma^{\mu \nu \xi})_a{}^{\dot b}\partial^\rho B_{\xi \mu} \partial_\nu \psi_{\rho \dot b}\\
    &\qquad \qquad+ (\sigma^{\mu \rho \xi})_{a}{}^{\dot b}\partial^\nu B_{\xi \mu} \partial_\nu\psi_{\rho \dot b} - (\sigma^{\nu \rho \xi})_a{}^{\dot b} \partial^\delta B_{\xi \delta} \partial_\nu \psi_{\rho \dot b}\biggr]\\
    &\quad+2b_3e_1\eta^{\textcolor{blue}{[\rho|} \delta} \eta^{\textcolor{blue}{|\nu|} \xi}(\sigma^{\textcolor{blue}{|\mu]}})_a{}^{\dot b}
     \partial_\delta B_{\xi \mu}    \partial_{\nu} \psi_{\rho \dot b}
\end{split}
\end{align}

\begin{align}
\begin{split}
    (C5) &= -\frac{b_3e_1}{4}(\sigma^{\mu \nu \rho} \tilde \sigma_\mu \sigma^{[3]})^{\dot b}{}_a A_{[3]}\partial_\nu \psi_{\rho \dot b}\\
    &= -2b_3e_1(\sigma^{\nu \rho} \sigma^{\delta \xi \mu})^{\dot b}{}_a \partial_\delta B_{\xi \mu}\partial_\nu \psi_{\rho \dot b}\\
    &= -2b_3e_1 \partial_\nu\biggl[(\sigma^{\mu \nu \rho \delta \xi})_{a}{}^{\dot b}\partial_\delta B_{\xi \mu} \psi_{\rho \dot b} + 2(\sigma^{\mu \nu \delta})_a{}^{\dot b}\partial_\delta B^\rho{}_{ \mu}\psi_{\rho \dot b}\biggr]\\
    &\quad -2b_3e_1\biggl[-2(\sigma^{\mu \rho \delta})_a{}^{\dot b}\partial_\delta B^\nu{}_\mu \partial_\nu\psi_{\rho \dot b} - (\sigma^{\mu \nu \xi})_a{}^{\dot b}\partial^\rho B_{\xi \mu} \partial_\nu \psi_{\rho \dot b}\\
    &\qquad \qquad +(\sigma^{\mu \rho \xi})_{a}{}^{\dot b}\partial^\nu B_{\xi \mu} \partial_\nu\psi_{\rho \dot b}\biggr]\\
    &\quad - 2b_3e_1\eta^{\rho\textcolor{blue}{[\delta|}}\eta^{\nu\textcolor{blue}{|\xi|}}(\sigma^{\textcolor{blue}{|\mu]}})_a{}^{\dot b} \partial_\delta B_{\xi \mu}    \partial_{\nu} \psi_{\rho \dot b}
\end{split}
\end{align}

\noindent Altogether for the gravitino we conclude
\begin{align}
\begin{split}
    ib_3 Q_a \big{[} \psi_{\mu}{}^c (\sigma^{\mu\nu\rho})_c{}^{\dot b} \partial_{\nu} \psi_{\rho \dot b} \big{]}
    &=b_3 \partial_\nu \biggl[-(\sigma^{\mu\nu  \delta})_a{}^{\dot b}\partial_\delta h^\rho_\mu \psi_{\rho \dot b}  -(\sigma^{\nu \rho \delta})_a{}^{\dot b}\partial_\delta h \psi_{\rho \dot b}\\
    &\qquad \qquad+  (\sigma^{\mu \rho \delta})_a{}^{\dot b}\partial_\delta h_{\mu }^\nu\psi_{\rho \dot b}+(\sigma^{\nu \rho \delta})_a{}^{\dot b}\partial^\mu h_{\mu \delta}\psi_{\rho \dot b}\biggr]\\
    &\quad -b_3\eta^{\textcolor{blue}{[\rho|} \delta} \eta^{\textcolor{blue}{|\nu|} \xi}(\sigma^{\textcolor{blue}{|\mu]}})_a{}^{\dot b}
     \partial_\delta h_{\xi \mu}    \partial_{\nu} \psi_{\rho \dot b}\\
     &\quad -2b_3e_1 \partial_\nu\biggl[ (\sigma^{\mu \nu \delta})_a{}^{\dot b}\partial_\delta B^\rho{}_{ \mu}\psi_{\rho \dot b}\biggr]\\
    &\quad +2b_3e_1\biggl[(\sigma^{\mu \rho \delta})_a{}^{\dot b}\partial_\delta B^\nu{}_\mu \partial_\nu\psi_{\rho \dot b}  - (\sigma^{\nu \rho \xi})_a{}^{\dot b} \partial^\delta B_{\xi \delta} \partial_\nu \psi_{\rho \dot b}\biggr]\\
    &\quad+2b_3e_1\eta^{\textcolor{blue}{[\rho|} \delta} \eta^{\textcolor{blue}{|\nu|} \xi}(\sigma^{\textcolor{blue}{|\mu]}})_a{}^{\dot b}
     \partial_\delta B_{\xi \mu}    \partial_{\nu} \psi_{\rho \dot b}\\
    &\quad - 2b_3e_1\eta^{\rho\textcolor{blue}{[\delta|}}\eta^{\nu\textcolor{blue}{|\xi|}}(\sigma^{\textcolor{blue}{|\mu]}})_a{}^{\dot b} \partial_\delta B_{\xi \mu}    \partial_{\nu} \psi_{\rho \dot b}\\
    &\quad  - ib_3 \partial_\nu \big{[} \psi_{\mu}{}^c (\sigma^{\mu\nu\rho})_c{}^{\dot b}  Q_a \psi_{\rho \dot b} \big{]}
\end{split}
\end{align}
Let's move on to the dilatino term.
\begin{align}
\begin{split}
    id_3 Q_a \left[\chi^{\dot c}(\tilde \sigma^\mu)_{\dot c}{}^b \partial_\mu \chi_b\right] &= 2id_3 Q_a \chi_c(\tilde \sigma^\mu)^{bc} \partial_\mu \chi_b - id_3 \partial_\nu \left[\chi^{\dot c}(\tilde \sigma^\nu)_{\dot c}{}^b Q_a\chi_b\right]\\
    &= -2d_3f_0( \sigma^\nu\tilde \sigma^\mu)_a{}^b \partial_\nu \phi\partial_\mu \chi_b + 2d_3e_0( \sigma^{[3]}\tilde \sigma^\mu)_a{}^b A_{[3]}\partial_\mu \chi_b\\
    &\quad- id_3 \partial_\nu \left[\chi^{\dot c}(\tilde \sigma^\nu)_{\dot c}{}^b Q_a\chi_b\right]\\
    &= -2d_3f_0\partial^\mu \phi \partial_\mu \chi_a -2d_3f_0 \partial_\nu \left[( \sigma^{\nu \mu})_a{}^b \phi \partial_\mu \chi_b\right]\\
    &\quad + 2d_3e_0\partial_\nu\left[( \sigma^{ [3]\nu})_a{}^b A_{[3]} \chi_b\right]- id_3 \partial_\nu \left[\chi^{\dot c}(\tilde \sigma^\nu)_{\dot c}{}^b Q_a\chi_b\right]\\
    &\quad + 2d_3e_0 \left[( \sigma^{\nu\rho})_a{}^b \partial^\mu B_{\nu \rho}\partial_\mu \chi_b + 2 ( \sigma^{\nu \rho})_a{}^b \partial_\nu B_{\rho \mu}\partial^\mu \chi_b\right]
\end{split}
\end{align}
We can now combine all of the above in order to determine our constraints. We omit the simplification of this equation, which consists of expanding commutators and matching cross terms.
\begin{align}
\begin{split}
    Q_a \mathcal L &= (b_3-2)(\sigma_{\mu})_a{}^{\dot b}\partial_{\rho} \psi_{\nu \dot b}\partial^{\rho}h^{\mu\nu} - (b_3-2) (\sigma^{\nu})_a{}^{\dot b}\partial^\rho \psi_{\nu \dot b}\partial_\rho h +(b_3-2)  (\sigma^{\nu})_a{}^{\dot b}\partial^\rho \psi_{\nu \dot b} \partial^\mu h_{\rho\mu}\\
    &\quad +(b_3-2) (\sigma_{\mu})_a{}^{\dot b} \partial^\rho  h \partial^\mu\psi_{\rho \dot b} + (2-b_3)  (\sigma_\mu)_a{}^{\dot b}\partial^{\mu}  \psi_{\nu \dot b} \partial_\rho h^{\rho \nu} +  (2-b_3)(\sigma_{\nu})_a{}^{\dot b}\partial^{\mu}  \psi_{\mu \dot b} \partial_\rho h^{\rho \nu} \\
     &\quad - (c_1+2d_3f_0) \partial_\mu \chi_a \partial^\mu \phi\\
    &\quad + (2d_3e_0 - 1)(\sigma^{\nu \rho})_a{}^b\partial^\mu B_{\nu \rho}\partial_\mu \chi_b + 2(2d_3e_0 - 1)(\sigma_{\mu \nu})_a{}^b \partial_\rho \chi_b \partial^\mu B^{\nu \rho} \\
    &\quad -(2a_2 + 2b_3e_1) (\sigma_\nu)_a{}^{\dot b} \partial_\mu \psi_{\rho \dot b} \partial^\mu B^{\nu \rho} - (2a_2 + 2b_3e_1)(\sigma_\mu)_{a}{}^{\dot b}\partial_\rho \psi_{\nu \dot b} \partial^\mu B^{\nu \rho} \\
    &\quad+ (2a_2 + 2b_3e_1)(\sigma_\nu)_a{}^{\dot b} \partial_\rho \psi_{\mu \dot b} \partial^\mu B^{\nu \rho}
    \\
    &\quad + 2b_3e_1 \partial_\nu \biggl[ (\sigma^\nu)_a{}^{\dot b} \partial^\mu B^\rho{}_\mu \psi_{\rho \dot b} - (\sigma^\mu)_a{}^{\dot b} \partial_\mu B^{\rho \nu} \psi_{\rho \dot b} - (\sigma^{\rho})_{a}{}^{\dot b} \partial_\mu B^{\nu \mu} \psi_{\rho \dot b}
     \\
     &\qquad\qquad \qquad +(\sigma^{\mu \rho \delta})_a{}^{\dot b}  \partial_\delta B^\nu{}_\mu \psi_{\rho \dot b} - (\sigma^{\mu \rho \nu})_a{}^{\dot b}\partial_\delta B^\delta{}_\mu \psi_{\rho \dot b}-(\sigma^{\mu \nu \delta})_a{}^{\dot b}\partial_\delta B^\rho{}_{ \mu}\psi_{\rho \dot b}\biggr]\\
    &\quad +\partial_\nu \left[(\sigma_{\mu})_a{}^{\dot b} \partial^\mu  h \psi^\nu{}_{ \dot b} - (\sigma^\nu)_a{}^{\dot b} \partial^\mu  h \psi_{\mu \dot b}+2(\sigma^\mu)_a{}^{\dot b} \partial^\rho h_{\rho \mu}\psi^{\nu}{}_{ \dot b} - 2(\sigma^\mu)_a{}^{\dot b} \partial^\rho h^\nu_{ \mu}\psi_{\rho \dot b} \right]\\
    &\quad +b_3 \partial_\nu \biggl[-(\sigma^{\mu\nu  \delta})_a{}^{\dot b}\partial_\delta h^\rho_\mu \psi_{\rho \dot b}  -(\sigma^{\nu \rho \delta})_a{}^{\dot b}\partial_\delta h \psi_{\rho \dot b}+  (\sigma^{\mu \rho \delta})_a{}^{\dot b}\partial_\delta h_{\mu }^\nu\psi_{\rho \dot b}\\
    &\qquad \qquad\quad+(\sigma^{\nu \rho \delta})_a{}^{\dot b}\partial^\mu h_{\mu \delta}\psi_{\rho \dot b}-i\psi_{\mu}{}^c (\sigma^{\mu\nu\rho})_c{}^{\dot b}  Q_a \psi_{\rho \dot b}\biggr]\\
    &\quad + 2d_3e_0\partial_\nu( \sigma^{ [3]\nu})_a{}^b A_{[3]} \chi_b - 2d_3f_0( \sigma^{ \nu\mu})_a{}^b \phi \partial_\mu \chi_b\\
    &\quad- id_3 \partial_\nu \left[\chi^{\dot c}(\tilde \sigma^\nu)_{\dot c}{}^b  Q_a\chi_b\right]
\end{split}
\end{align}
In order for our supercurrent to assume the proper form, we require that $b_3 = 2$, $c_1 = -2d_3f_0$, $2d_3e_0 = 1$, $a_2 = -2e_1$. Recall that the constraints resulting from closure of the superalgebra are $d_0 = -1/2$, $a_2e_1 = -1/2$, $c_1f_0 = 1$, $e_2 = -e_1/8$, and $e_0 = -1/8$. These are 9 independent constraints on 9 independent coefficients (7 from the superalgebra, 2 from the Lagrangian density) which have the following unique solution on the superalgebra coefficients (up to freedom of some signs):
\begin{equation}
\begin{alignedat}{3}
a_1&=1            &\qquad c_1&=\sqrt{8}            &\qquad d_0&=-\tfrac12\\
a_2&=1            &       f_0&=\tfrac{1}{\sqrt{8}}  &       e_1&=-\tfrac12\\
c_2&=1            &       e_0&=-\tfrac{1}{8}       &       e_2&=\tfrac{1}{16}
\end{alignedat}
\end{equation}
and on the Lagrangian coefficients:
\begin{equation}
b_3 = 2 \qquad \qquad d_3 = -4
\end{equation}
We conclude that
\begin{align}
\begin{split}
    Q_a \mathcal L = \partial_\nu (J^\nu)_a = \partial_\nu &\Biggl\{
         -2(\sigma^\nu)_a{}^{\dot b} \partial^\mu B^\rho{}_\mu \psi_{\rho \dot b} +2 (\sigma^\mu)_a{}^{\dot b} \partial_\mu B^{\rho \nu} \psi_{\rho \dot b} +2 (\sigma^{\rho})_{a}{}^{\dot b} \partial_\mu B^{\nu \mu} \psi_{\rho \dot b}
     \\
     &\quad -2(\sigma^{\mu \rho \delta})_a{}^{\dot b}  \partial_\delta B^\nu{}_\mu \psi_{\rho \dot b} +2 (\sigma^{\mu \rho \nu})_a{}^{\dot b}\partial_\delta B^\delta{}_\mu \psi_{\rho \dot b}+2(\sigma^{\mu \nu \delta})_a{}^{\dot b}\partial_\delta B^\rho{}_{ \mu}\psi_{\rho \dot b}\\
    &\quad +(\sigma_{\mu})_a{}^{\dot b} \partial^\mu  h \psi^\nu{}_{ \dot b} - (\sigma^\nu)_a{}^{\dot b} \partial^\mu  h \psi_{\mu \dot b}+2(\sigma^\mu)_a{}^{\dot b} \partial^\rho h_{\rho \mu}\psi^{\nu}{}_{ \dot b} - 2(\sigma^\mu)_a{}^{\dot b} \partial^\rho h^\nu_{ \mu}\psi_{\rho \dot b} \\
    &\quad -2(\sigma^{\mu\nu  \delta})_a{}^{\dot b}\partial_\delta h^\rho_\mu \psi_{\rho \dot b}  -2(\sigma^{\nu \rho \delta})_a{}^{\dot b}\partial_\delta h \psi_{\rho \dot b}+  2(\sigma^{\mu \rho \delta})_a{}^{\dot b}\partial_\delta h_{\mu }^\nu\psi_{\rho \dot b}\\
    &\quad+2(\sigma^{\nu \rho \delta})_a{}^{\dot b}\partial^\mu h_{\mu \delta}\psi_{\rho \dot b}-2i\psi_{\mu}{}^c (\sigma^{\mu\nu\rho})_c{}^{\dot b}  Q_a \psi_{\rho \dot b}\\
    &\quad +( \sigma^{ [3]\nu})_a{}^b A_{[3]} \chi_b +\sqrt 8 ( \sigma^{ \nu\mu})_a{}^b \phi \partial_\mu \chi_b+4i \chi^{\dot c}(\tilde \sigma^\nu)_{\dot c}{}^b  Q_a\chi_b
    \Biggr\}
\end{split}
\end{align}
One can expand $4i \chi^{\dot c}(\tilde \sigma^\nu)_{\dot c}{}^b  Q_a\chi_b$ and $-2i\psi_{\mu}{}^c (\sigma^{\mu\nu\rho})_c{}^{\dot b}  Q_a \psi_{\rho \dot b}$ to arrive at
\begin{align}
\begin{split}
    Q_a \mathcal L = \partial_\nu (J^\nu)_a = \partial_\nu &\Biggl\{
         -(\sigma^\nu)_a{}^{\dot b} \partial^\mu B^\rho{}_\mu \psi_{\rho \dot b}  + (\sigma^{\rho})_{a}{}^{\dot b} \partial_\mu B^{\nu \mu} \psi_{\rho \dot b} \\
    &\quad + (\sigma^{\rho})_{a c}  \partial^\nu B^\mu{}_{\rho } \psi_\mu{}^c -(\sigma^{\rho})_{a c}   \partial^\mu B^\nu{}_\rho \psi_\mu{}^c  - (\sigma^{\mu})_{a c}   \partial^\rho B^\nu{}_\rho \psi_\mu{}^c  \\
    &\quad -(\sigma^{\mu \rho \delta})_a{}^{\dot b}  \partial_\delta B^\nu{}_\mu \psi_{\rho \dot b} + (\sigma^{\mu \rho \nu})_a{}^{\dot b}\partial_\delta B^\delta{}_\mu \psi_{\rho \dot b}+(\sigma^{\mu \nu \delta})_a{}^{\dot b}\partial_\delta B^\rho{}_{ \mu}\psi_{\rho \dot b} \\
    &\quad -(\sigma^{\mu})_{c a}   \partial^\rho h^\nu{}_\rho \psi_\mu{}^c + (\sigma^{\mu})_{c a}  \partial^\nu h \psi_\mu{}^c  -(\sigma^{\rho})_{c a}  \partial^\nu h^\mu{}_\rho \psi_\mu{}^c \\
    &\quad -(\sigma^\mu)_a{}^{\dot b} \partial^\rho h^\nu{}_{ \mu}\psi_{\rho \dot b}-2 (\sigma^\nu)_a{}^{\dot b} \partial^\mu  h \psi_{\mu \dot b} + \psi_\mu{}^c (\sigma^{\nu})_{c a} \partial_\delta h^{\mu \delta} \\
    &\quad +(\sigma_{\mu})_a{}^{\dot b} \partial^\mu  h \psi^\nu{}_{ \dot b} +2(\sigma^\mu)_a{}^{\dot b} \partial^\rho h_{\rho \mu}\psi^{\nu}{}_{ \dot b} \\
    &\quad -(\sigma^{\mu\nu  \delta})_a{}^{\dot b}\partial_\delta h^\rho_\mu \psi_{\rho \dot b}  -(\sigma^{\nu \rho \delta})_a{}^{\dot b}\partial_\delta h \psi_{\rho \dot b}+  (\sigma^{\mu \rho \delta})_a{}^{\dot b}\partial_\delta h_{\mu }^\nu\psi_{\rho \dot b}\\
    &\quad+(\sigma^{\nu \rho \delta})_a{}^{\dot b}\partial^\mu h_{\mu \delta}\psi_{\rho \dot b} +\frac{1}{2}( \sigma^{ [3]\nu})_a{}^b A_{[3]} \chi_b  +\sqrt 2 ( \sigma^{ \nu\mu})_a{}^b \phi \partial_\mu \chi_b \\
    &\quad -\sqrt 2  \partial^\nu \phi \chi_a  + \frac{1}{2}   (\sigma^{ \rho \delta})_{\dot c a}  \partial^\nu B_{\rho \delta}\chi^{\dot c} +    (\sigma^{ \mu \rho})_{\dot c a} \partial_\mu B_{\rho }{}^\nu \chi^{\dot c}
    \Biggr\}
\end{split}
\end{align}

It is then easily computed that
\begin{align}
\begin{split}
    \sum_i \frac{\partial \mathcal L}{\partial (\partial_\nu \Phi_i)}\Phi_i &=-\sqrt 8\partial^\nu \phi \chi_a -6A^{\nu \mu \rho}(\sigma_{\mu})_a{}^{\dot b}\psi_{\rho \dot b} -3A^{\nu \mu \rho}(\sigma_{\mu \rho})_a{}^b \chi_b\\
    &\quad +\sqrt 2 \chi^{\dot c}(\sigma^\rho\tilde \sigma^\nu)_{a \dot c } \partial_\rho \phi + \frac 12 \chi^{\dot c}( \sigma^{[3]}\tilde \sigma^{\nu})_{a \dot c}A_{[3]}\\
    &\quad+\psi_{\mu}{}^c(\sigma^{\mu \nu \rho}\tilde \sigma^{\lambda \delta})_{ca}\partial_\lambda h_{\delta \rho}+\psi_{\mu}{}^c(\sigma^{\mu \nu \rho}\tilde \sigma^{\lambda \delta})_{ca}\partial_\lambda B_{\delta \rho}- \frac 18 \psi_{\mu}{}^c (\sigma^{\mu \nu \rho}\tilde \sigma_\rho \sigma^{[3]})_{ca}A_{[3]}\\
    &\quad-2\partial^\nu h^{\mu \rho} \left(\sigma_{\mu}\right)_a{}^{\dot b}\psi_{\rho \dot b} +2\partial^\nu h \left(\sigma^{\mu}\right)_a{}^{\dot b}\psi_{\mu \dot b}-2\partial_\beta h^{\nu \beta} \left(\sigma^{\mu}\right)_a{}^{\dot b}\psi_{\mu \dot b}\\&\quad-\partial_{\mu} h (\sigma^{(\nu})_a{}^{\dot b}\psi^{\mu)}{}_{ \dot b}+2\partial^\alpha h_{\alpha \mu}(\sigma^{(\nu})_a{}^{\dot b}\psi^{\mu)}{}_{\dot b}
\end{split}
\end{align}
Expanding these terms results in the following expression:
\begin{align}
\begin{split}
    \sum_i \frac{\partial \mathcal L}{\partial (\partial_\nu \Phi_i)}\Phi_i &=(\sqrt 2-\sqrt 8)\partial^\nu \phi \chi_a -\frac 3 2 A^{\nu \mu \rho}(\sigma_{\mu \rho})_a{}^b \chi_b-\sqrt 2 \chi^{\dot c}( \sigma^{\nu \rho})_{a \dot c} \partial_\rho \phi + \frac 12 \chi^{\dot c}(  \sigma^{[3]\nu})_{a\dot c}A_{[3]}\\
    &\quad+\psi_{\mu}{}^c(\sigma^{\mu \nu \delta})_{ca}\partial^\rho h_{\delta \rho}-\psi_{\mu}{}^c(\sigma^{\mu \nu \lambda})_{ca}\partial_\lambda h + \psi_{\mu}{}^c(\sigma^{\mu \rho \lambda})_{ca}\partial_\lambda h^\nu{}_\rho-\psi_{\mu}{}^c(\sigma^{\nu \rho \lambda})_{ca}\partial_\lambda h^\mu{}_\rho\\
    &\quad + \psi_\lambda{}^c (\sigma^\nu)_{ca} \partial^\lambda h -  \psi_\mu{}^c (\sigma^\mu)_{ca} \partial^\nu h +  \psi_\mu{}^c (\sigma^\mu)_{ca} \partial^\rho h^\nu{}_\rho -  \psi_\mu{}^c (\sigma^\rho)_{ca} \partial^\mu h^\nu{}_\rho\\
    &\quad + \psi_\mu{}^c (\sigma^\rho)_{ca} \partial^\nu h^\mu{}_\rho -  \psi_\mu{}^c (\sigma^\nu)_{ca} \partial^\rho h^\mu{}_\rho\\
    &\quad+\psi_{\mu}{}^c(\sigma^{\mu \nu \delta})_{ca}\partial^\rho B_{\delta \rho}+\psi_{\mu}{}^c(\sigma^{\mu \rho \lambda})_{ca}\partial_\lambda B^{\nu}{}_{ \rho}-\psi^{\mu c}(\sigma^{ \nu \rho \lambda})_{ca}\partial_\lambda B_{\mu \rho}\\
    &\quad + \psi_{\mu}{}^c(\sigma^{\nu \rho \delta})_{ca}\partial^\mu B_{\delta \rho}-\psi_{\mu}{}^c(\sigma^{\mu \rho \delta })_{ca}\partial^\nu B_{\delta \rho}\\
     &\quad +  \psi_\mu{}^c (\sigma^\mu)_{ca} \partial^\rho B^\nu{}_\rho -  \psi_\mu{}^c (\sigma^\rho)_{ca} \partial^\mu B^\nu{}_\rho+ \psi_\mu{}^c (\sigma^\rho)_{ca} \partial^\nu B^\mu{}_\rho -  \psi_\mu{}^c (\sigma^\nu)_{ca} \partial^\rho B^\mu{}_\rho\\
    &\quad - 3\psi_{\mu}{}^c (\sigma^{[\mu }{}_{[2]})_{ca}A^{\nu][2]} \\
    &\quad-2\partial^\nu h^{\mu \rho} \left(\sigma_{\mu}\right)_a{}^{\dot b}\psi_{\rho \dot b} +2\partial^\nu h \left(\sigma^{\mu}\right)_a{}^{\dot b}\psi_{\mu \dot b}-2\partial_\beta h^{\nu \beta} \left(\sigma^{\mu}\right)_a{}^{\dot b}\psi_{\mu \dot b}\\&\quad-\partial_{\mu} h (\sigma^{(\nu})_a{}^{\dot b}\psi^{\mu)}{}_{ \dot b}+2\partial^\alpha h_{\alpha \mu}(\sigma^{(\nu})_a{}^{\dot b}\psi^{\mu)}{}_{\dot b}
\end{split}
\end{align}
Combining Eqs. (C.16) and (C.18) results in the final supercurrent:
\begin{align}
\begin{split}
    (J^\mu)_a &=
     -3 A^{\nu \mu \rho}(\sigma_{\mu \rho})_a{}^b \chi_b - 2\sqrt 2 ( \sigma^{ \nu\mu})_a{}^b \phi \partial_\mu \chi_b \\
    &\quad + 3\psi_\lambda{}^c (\sigma^\nu)_{ca} \partial^\lambda h -  2\psi_\mu{}^c (\sigma^\mu)_{ca} \partial^\nu h +  2\psi_\mu{}^c (\sigma^\mu)_{ca} \partial^\rho h^\nu{}_\rho \\
    &\quad +2 \psi_\mu{}^c (\sigma^\rho)_{ca} \partial^\nu h^\mu{}_\rho -  2\psi_\mu{}^c (\sigma^\nu)_{ca} \partial^\rho h^\mu{}_\rho\\
    &\quad -(\sigma_{\mu})_{ca} \partial^\mu  h \psi^{\nu c} -2(\sigma^\mu)_{ca} \partial^\rho h_{\rho \mu}\psi^{\nu c}\\
    &\quad+2\psi_{\mu}{}^c(\sigma^{\mu \rho \lambda})_{ca}\partial_\lambda B^{\nu}{}_{ \rho}-2\psi^{\mu c}(\sigma^{ \nu \rho \lambda})_{ca}\partial_\lambda B_{\mu \rho}\\
    &\quad + \psi_{\mu}{}^c(\sigma^{\nu \rho \delta})_{ca}\partial^\mu B_{\delta \rho}-\psi_{\mu}{}^c(\sigma^{\mu \rho \delta })_{ca}\partial^\nu B_{\delta \rho}\\
    &\quad +\psi_\mu{}^c (\sigma^{\mu})_{ca}   \partial_\rho B^{\nu \rho} - 3\psi_{\mu}{}^c (\sigma^{[\mu }{}_{[2]})_{ca}A^{\nu][2]} \\
    &\quad-2\partial^\nu h^{\mu \rho} \left(\sigma_{\mu}\right)_a{}^{\dot b}\psi_{\rho \dot b} +2\partial^\nu h \left(\sigma^{\mu}\right)_a{}^{\dot b}\psi_{\mu \dot b}-2\partial_\beta h^{\nu \beta} \left(\sigma^{\mu}\right)_a{}^{\dot b}\psi_{\mu \dot b}\\&\quad-\partial_{\mu} h (\sigma^{(\nu})_a{}^{\dot b}\psi^{\mu)}{}_{ \dot b}+2\partial^\alpha h_{\alpha \mu}(\sigma^{(\nu})_a{}^{\dot b}\psi^{\mu)}{}_{\dot b}
\end{split}
\end{align}

{\bf Acknowledgment.}
The research of S.J.G. is currently supported in part by the Clark Leadership Chair in Science endowment at the University of Maryland - College Park and the Army Research Office under Contract WNW911NF2520117. The authors are grateful to Dr. Konstantinos Koutrolikos for his mentorship and guidance on this project, including regular discussions, technical oversight, and close supervision of the work of Bergen Dahl and Jacob Cigliano. B.D. and J.C. also acknowledge their participation in the Summer Student Theoretical Physics Research Session (SSTPRS) during June of 2024.

\bibliographystyle{hephys} 
\bibliography{refs_fixed}

\end{document}